\documentclass[twocolumn,longbibliography,superscriptaddress,
showpacs,preprintnumbers,
notitlepage,amsmath,amssymb,
aps,
]{revtex4-1}

\usepackage{graphicx}
\usepackage{dcolumn}
\usepackage{bm}
\usepackage{bbm}
\usepackage{diagbox}

\usepackage{graphicx}
\usepackage{enumerate}
\usepackage{subfigure}
\usepackage{epstopdf}
\usepackage[colorlinks,linkcolor=red,anchorcolor=black,citecolor=blue]{hyperref}
\usepackage{amssymb}
\usepackage{framed}
\usepackage{tabularx}
\usepackage{booktabs}
\usepackage{float}
\usepackage[table]{xcolor}
\usepackage{array}
\usepackage{tikz}
\hypersetup{colorlinks=true}

\newcommand{\bs}{\boldsymbol}

\newcommand{\1}{\text{\uppercase\expandafter{\romannumeral1}}}
\newcommand{\2}{\text{\uppercase\expandafter{\romannumeral2}}}
\newcommand{\3}{\text{\uppercase\expandafter{\romannumeral3}}}
\newcommand{\4}{\text{\uppercase\expandafter{\romannumeral4}}}
\newcommand{\5}{\text{\uppercase\expandafter{\romannumeral5}}}
\newcommand{\6}{\text{\uppercase\expandafter{\romannumeral6}}}

\def\U{\mathrm{U}(1)}
\def\H{\mathcal{H}}
\def\Z{\mathbb{Z}}

\begin{document}
\title{Classification and construction of interacting fractonic higher-order topological phases}
\author{Jian-Hao Zhang}
\affiliation{Department of Physics, The Pennsylvania State University, University Park, Pennsylvania 16802, USA}
\author{Meng Cheng}
\email{m.cheng@yale.edu}
\affiliation{Department of Physics, Yale University, New Haven, Connecticut 06511-8499, USA}
\author{Zhen Bi}
\email{zjb5184@psu.edu}
\affiliation{Department of Physics, The Pennsylvania State University, University Park, Pennsylvania 16802, USA}

\begin{abstract}
The notion of higher-order topological phases can have interesting generalizations to systems with subsystem symmetries that exhibit fractonic dynamics for charged excitations. In this work, we systematically study the higher-order topological phases protected by a combination of subsystem symmetries and ordinary global symmetries in two and three-dimensional interacting boson systems, with some interacting fermionic examples. 
\end{abstract}

\maketitle
\tableofcontents

\section{Introduction}
Symmetry-protected topological (SPT) phases greatly expand our knowledge of quantum phases of matter beyond the conventional Landau symmetry-breaking paradigm \cite{ZCGu2009, XieChenScience, cohomology, Senthil_2015}. Tremendous progress has been made over the past decade in classifying and characterizing SPT phases with internal and crystalline symmetries \cite{Lu12,invertible2,invertible3, special,general1,general2,Kapustin2014, Kapustin2015,Kapustin2017,gauging2,2DFSPT}. A common feature of SPT phases is that the boundary of an SPT phase is usually gapless due to symmetry protection. For instance, the celebrated $3d$ topological insulator hosts  single massless Dirac cones protected by charge conservation and time reversal symmetry on its $2d$ surfaces. However, in contrast to the phenomenology of the ordinary topological insulator, recently a new class of SPT phases is shown to exist where symmetry-protected gapless modes only show up on certain low-dimensional submanifolds on the boundary while the majority of the boundary can be gapped without breaking the symmetry. These features defined a new class of SPT phases which is dubbed higher-order topological phases \cite{higher1,higher2,higher3,higher4,higher5}.

Higher-order topological phases turn out to be rather common in systems with crystalline symmetries. A great deal of weakly interacting higher-order topological insulators and superconductors has been established theoretically \cite{TCI,Fu2012,ITCI,reduction,building,correspondence,indicator1,bultinck2019three,Roy_2020,Roy_2021,Laubscher_2019,Laubscher_2020,Zhang_2022} and discovered experimentally \cite{TCIrealization1,TCIrealization2,TCIrealization3,TCIrealization4} as well. For strongly interacting systems of fermions or bosons, one can also demonstrate the existence and study the properties of higher-order topological phases with methods such as the crystalline equivalence principle \cite{correspondence} and the block state constructions \cite{ZDSong2018, realspace, rotation, dihedral, wallpaper, JHZhang2022}. These studies bring us a complete picture of symmetry-protected topological phases with crystalline symmetries.  

Coming from a rather orthogonal direction, a new type of symmetry, namely subsystem symmetry, is discovered along the exploration of so-called fracton topological phases. Compared to ordinary global symmetry, subsystem symmetries are more closely intertwined with the underlying foliation \cite{Vijay_2016, Nandkishore_2019, Pretko_2020, Pai_2019} structure of space and only act on a rigid submanifold/leaf of the whole system. As the charges on each submanifold are conserved individually, single charge tunneling events are forbidden, which leads to fractonic behaviors \cite{Vijay_2016, Nandkishore_2019, Pretko_2020, Pai_2019}. Approximate subsystem symmetries have natural manifestations in certain physical systems. For instance, in twisted bilayer tungsten ditelluride systems, at a twist angle of $5^\circ$, experiments \cite{quasi1d} exhibit exceptionally large transport anisotropy between two orthogonal in-plane directions, indicating vanishing single-particle tunneling in one direction. Another system with approximate subsystem symmetry is $3d$ kagome metal materials, Ni$_3$In as an example \cite{ye2021flat}, where the single-particle tunnelings within $xy$-plane are suppressed due to interference effects of atomic orbitals. Subsystem symmetry may also be an emergent symmetry at some quantum critical points \cite{decoupling}.

A natural question is whether there are nontrivial SPT phases associated with subsystem symmetry. Indeed, previous works have shown the existence of subsystem symmetry-protected topological (SSPT) phases \cite{You_2018, Devakul_2019, Devakul_2018, Devakul_2020, Williamson_2019, May_Mann_2019, You_2020, Stephen_2020, May_Mann_2021}. Examples of higher-order topological phases with subsystem symmetries are also discovered \cite{May-Mann2022} but still lack a systematic understanding. In this work, we systematically study possible higher-order topological phases in strongly interacting bosonic systems protected by a combination of global and subsystem symmetries labeled by $G_g\times G_s$. We present a general scheme for classifying higher-order SSPT states, that applies to both bosonic and fermionic systems and work out the complete mathematical classification in the bosonic case. From the general classification, we establish a few interesting facts. For instance, for Abelian subsystem symmetry, there is no nontrivial 2-foliated higher-order SSPT phase in (2+1)d systems without the aid of global symmetry. In addition, we prove that for inhomogeneous subsystem symmetries, there is no nontrivial higher-order SSPT phase. Besides the general classification, we also explicitly construct models of such SSPT states in 2 and 3 spatial dimensions.

The rest of the paper is organized as follows. In Sec. \ref{Sec. 2-foliated}, we discuss the classifications and explicit model constructions of higher-order SPT phases with 2-foliated subsystem symmetry.  In particular, we prove that if $G_g$ is trivial, there is no nontrivial higher-order SSPT phase in (2+1)D systems.  Finally, in Sec. \ref{Sec.corner}, we consider 3-dimensional systems with the 3-foliation structure, including the classifications and explicit lattice model constructions.  We will explicitly construct an exactly solvable lattice model of (3+1)D third-order SSPT phase with 3-foliated $\mathbb{Z}_2\times\Z_2$ subsystem symmetry. Furthermore, we will demonstrate that for arbitrary foliated inhomogeneous subsystem symmetries in any dimension, there is no nontrivial higher-order SSPT phase. In Sec. \ref{Sec.summary}, we summarize the main results of this paper and discuss further outlooks.

\section{Second-order SSPT phases with 2-foliated subsystem symmetry\label{Sec. 2-foliated}}

A foliation is a decomposition of a manifold into an infinite number of disjoint lower-dimensional submanifolds called leaves. By 2-foliation, we mean the physical system we consider can be decomposed into two orthogonal sets of disjoint codimension-1 submanifolds. And an independent conserved symmetry charge can be defined on each of these codimension-1 subsystems. 

Before passing to the 2-foliated systems, we want to mention that 1-foliated systems cannot host any nontrivial higher-order SSPT phases even with global symmetry. The argument is presented in Appendix. \ref{1foliated}.

In this section, we consider the system with 2-foliated subsystem symmetries. We will assume that the symmetries are on-site, so the unitaries that implement the symmetry transformations are tensor products of unitaries acting on the Hilbert spaces of sites. We will generally consider the following two scenarios: 1) homogeneous subsystem symmetries where the 2-foliated subsystem symmetries are built out of the same on-site symmetry transformations. 2) inhomogeneous subsystem symmetries where the symmetries in different directions act completely differently. As we will see, the two kinds of subsystem symmetries have significant physical differences, therefore, we discuss them separately. Schematically, we denote the subsystem symmetry group by $G_s$. In addition to the subsystem symmetries, we also consider an additional global symmetry, $G_g$, in the system. For simplicity, in this work, we consider the situation where the total symmetry is a direct product of $G_s$ and $G_g$, although our formalism can be generalized to situations where the two groups have nontrivial central extensions straightforwardly.

\subsection{General remarks}

For a $(d+1)$D lattice with a 2-foliation structure, for our purpose, it is convenient to think of the system as a square grid. Each site-$(x,y)$ of the grid corresponds to a $(d-2)$ system, whose Hilbert space is denoted by $\mathcal{H}_{xy}$. For $d>2$ the $(d-2)$D system itself may be extensive. The total Hilbert space is a tensor product $\mathcal{H}=\otimes_{x,y}\mathcal{H}_{xy}$. The 2-foliated homogeneous subsystem symmetries are defined as follows,
\begin{align}
\begin{aligned}
&U_x(g)=\prod\limits_{y=-\infty}^{\infty}u_{xy}(g)\\
&U_y(g)=\prod\limits_{x=-\infty}^\infty u_{xy}(g)
\end{aligned}~,~~g\in G_s
\label{(2+1)DSS}
\end{align}
where $u_{xy}(g)$ is an on-site unitary operator acting on the $(d-2)$D system at site $(x,y)$, and forms a faithful linear representation of the subsystem symmetry group $G_s$. The geometry of subsystem symmetries in $(2+1)$D systems is illustrated in Fig. \ref{2dSSym}. 

So far $G_s$ may be Abelian or non-Abelian.  Suppose $G_s$ is a non-Abelian group. For a given site $(x,y)$, and $g_1,g_2\in G_s$ we consider the following unitary:
\begin{align}
U_y(g_2^{-1})U_{x}(g_1^{-1})U_y(g_2)U_x(g_1)=u_{xy}(g_2^{-1}g_1^{-1}g_2g_1).
\end{align}
This operator is nontrivial if and only if $g_2^{-1}g_1^{-1}g_2g_1\ne1$. This commutator becomes a local symmetry of the site when this is the case. More generally, there is a local symmetry group $[G_s, G_s]$ on each site as the commutator of the subsystem symmetry $G_s$, and an effective Abelian subsystem symmetry $G_s'=G_s/[G_s, G_s]$ known as the Abelianization of $G_s$. Therefore, it is sufficient to consider the Abelian subgroup of the non-Abelian subsystem symmetry \cite{Devakul_2018}.

There is another possible pattern of subsystem symmetry: the subsystem symmetries along the two directions are completely different, i.e., they have different group structures and have different actions. Suppose the subsystem symmetry along the horizontal/vertical direction is $G_s^x/G_s^y$,  the 2-foliated subsystem symmetries should be defined as ($\forall x,y\in\mathbb{Z}$):
\begin{align}
\begin{aligned}
&U_x(g^y)=\prod\limits_{y=-\infty}^\infty u_{xy}^y(g^y)\\
&U_y(g^x)=\prod\limits_{x=-\infty}^{\infty}u_{xy}^x(g^x)
\end{aligned}~,~~\left\{
\begin{aligned}
&g^x\in G_s^x\\
&g^y\in G_s^y
\end{aligned}
\right.
\label{2-foliated inhomogeneous}
\end{align}
where $u_{xy}^y(g^y)/u_{xy}^x(g^x)$ is the linear representation of $G_s^x/G_s^y$ on the site-$(x,y)$, and $u_{xy}(g)=u_{xy}^y(g^y)u_{xy}^x(g^x)$ is the linear representation of $G_s^y\times G_s^x$ on the site-$(x,y)$, $g=g^xg^y$. 

We first consider systems with homogeneous subsystem symmetries and derive a complete classification. 

\subsection{Classification using boundary anomaly\label{2-foliated classification}}
\begin{figure}
\begin{tikzpicture}[scale=0.88]
\tikzstyle{sergio}=[rectangle,draw=none]
\path (-3.5,0.75) node [style=sergio] {$\hat{y}$};
\path (-2.25,2) node [style=sergio] {$\hat{x}$};
\draw[thick,->] (-3.5,2) -- (-2.5,2);
\draw[thick,->] (-3.5,2) -- (-3.5,1);
\filldraw[fill=blue!20, draw=blue, thick] (-2.75,1.25)--(3.75,1.25)--(3.75,0.75)--(-2.75,0.75)--cycle;
\filldraw[fill=blue!20, draw=blue, thick] (2.75,-4.75)--(2.75,1.75)--(3.25,1.75)--(3.25,-4.75)--cycle;
\path (3,2.1) node [style=sergio] {$U_{x}(g^y)$};
\path (4.4,1) node [style=sergio] {$U_{y}(g^x)$};
\draw[thick] (-2.5,1) -- (3.5,1);
\draw[ultra thick, color=red] (-2.5,0) -- (3.5,0);
\draw[thick] (-2.5,0.5) -- (3.5,0.5);
\draw[thick] (-2.5,-1) -- (3.5,-1);
\draw[thick] (-2.5,-0.5) -- (3.5,-0.5);
\draw[thick] (-2.5,-2) -- (3.5,-2);
\draw[thick] (-2.5,-1.5) -- (3.5,-1.5);
\draw[ultra thick, color=red] (-2.5,-3) -- (3.5,-3);
\draw[thick] (-2.5,-2.5) -- (3.5,-2.5);
\draw[thick] (-2.5,-4) -- (3.5,-4);
\draw[thick] (-2.5,-3.5) -- (3.5,-3.5);
\draw[thick] (-2,1.5) -- (-2,-4.5);
\draw[ultra thick, color=red] (-1.5,1.5) -- (-1.5,-4.5);
\draw[thick] (-1,1.5) -- (-1,-4.5);
\draw[thick] (-0.5,1.5) -- (-0.5,-4.5);
\draw[thick] (0,1.5) -- (0,-4.5);
\draw[thick] (0.5,1.5) -- (0.5,-4.5);
\draw[thick] (1,1.5) -- (1,-4.5);
\draw[ultra thick, color=red] (1.5,1.5) -- (1.5,-4.5);
\draw[thick] (2,1.5) -- (2,-4.5);
\draw[thick] (2.5,1.5) -- (2.5,-4.5);
\draw[thick] (3,1.5) -- (3,-4.5);
\filldraw[fill=black, draw=black] (0,1)circle (2.5pt);
\filldraw[fill=black, draw=black] (0.5,1)circle (2.5pt);
\filldraw[fill=black, draw=black] (1,1)circle (2.5pt);
\filldraw[fill=black, draw=black] (1.5,1)circle (2.5pt);
\filldraw[fill=black, draw=black] (-2,1)circle (2.5pt);
\filldraw[fill=black, draw=black] (-1.5,1)circle (2.5pt);
\filldraw[fill=black, draw=black] (-1,1)circle (2.5pt);
\filldraw[fill=black, draw=black] (-0.5,1)circle (2.5pt);
\filldraw[fill=black, draw=black] (2,1)circle (2.5pt);
\filldraw[fill=black, draw=black] (2.5,1)circle (2.5pt);
\filldraw[fill=black, draw=black] (3,1)circle (2.5pt);
\filldraw[fill=black, draw=black] (0,0.5)circle (2.5pt);
\filldraw[fill=black, draw=black] (0.5,0.5)circle (2.5pt);
\filldraw[fill=black, draw=black] (1,0.5)circle (2.5pt);
\filldraw[fill=black, draw=black] (1.5,0.5)circle (2.5pt);
\filldraw[fill=black, draw=black] (-2,0.5)circle (2.5pt);
\filldraw[fill=black, draw=black] (-1.5,0.5)circle (2.5pt);
\filldraw[fill=black, draw=black] (-1,0.5)circle (2.5pt);
\filldraw[fill=black, draw=black] (-0.5,0.5)circle (2.5pt);
\filldraw[fill=black, draw=black] (2,0.5)circle (2.5pt);
\filldraw[fill=black, draw=black] (2.5,0.5)circle (2.5pt);
\filldraw[fill=black, draw=black] (3,0.5)circle (2.5pt);
\filldraw[fill=black, draw=black] (0,0)circle (2.5pt);
\filldraw[fill=black, draw=black] (0.5,0)circle (2.5pt);
\filldraw[fill=black, draw=black] (1,0)circle (2.5pt);
\filldraw[fill=black, draw=black] (1.5,0)circle (2.5pt);
\filldraw[fill=black, draw=black] (-2,0)circle (2.5pt);
\filldraw[fill=black, draw=black] (-1.5,0)circle (2.5pt);
\filldraw[fill=black, draw=black] (-1,0)circle (2.5pt);
\filldraw[fill=black, draw=black] (-0.5,0)circle (2.5pt);
\filldraw[fill=black, draw=black] (2,0)circle (2.5pt);
\filldraw[fill=black, draw=black] (2.5,0)circle (2.5pt);
\filldraw[fill=black, draw=black] (3,0)circle (2.5pt);
\filldraw[fill=black, draw=black] (0,-0.5)circle (2.5pt);
\filldraw[fill=black, draw=black] (0.5,-0.5)circle (2.5pt);
\filldraw[fill=black, draw=black] (1,-0.5)circle (2.5pt);
\filldraw[fill=black, draw=black] (1.5,-0.5)circle (2.5pt);
\filldraw[fill=black, draw=black] (-2,-0.5)circle (2.5pt);
\filldraw[fill=black, draw=black] (-1.5,-0.5)circle (2.5pt);
\filldraw[fill=black, draw=black] (-1,-0.5)circle (2.5pt);
\filldraw[fill=black, draw=black] (-0.5,-0.5)circle (2.5pt);
\filldraw[fill=black, draw=black] (2,-0.5)circle (2.5pt);
\filldraw[fill=black, draw=black] (2.5,-0.5)circle (2.5pt);
\filldraw[fill=black, draw=black] (3,-0.5)circle (2.5pt);
\filldraw[fill=black, draw=black] (0,-1)circle (2.5pt);
\filldraw[fill=black, draw=black] (0.5,-1)circle (2.5pt);
\filldraw[fill=black, draw=black] (1,-1)circle (2.5pt);
\filldraw[fill=black, draw=black] (1.5,-1)circle (2.5pt);
\filldraw[fill=black, draw=black] (-2,-1)circle (2.5pt);
\filldraw[fill=black, draw=black] (-1.5,-1)circle (2.5pt);
\filldraw[fill=black, draw=black] (-1,-1)circle (2.5pt);
\filldraw[fill=black, draw=black] (-0.5,-1)circle (2.5pt);
\filldraw[fill=black, draw=black] (2,-1)circle (2.5pt);
\filldraw[fill=black, draw=black] (2.5,-1)circle (2.5pt);
\filldraw[fill=black, draw=black] (3,-1)circle (2.5pt);
\filldraw[fill=black, draw=black] (0,-1.5)circle (2.5pt);
\filldraw[fill=black, draw=black] (0.5,-1.5)circle (2.5pt);
\filldraw[fill=black, draw=black] (1,-1.5)circle (2.5pt);
\filldraw[fill=black, draw=black] (1.5,-1.5)circle (2.5pt);
\filldraw[fill=black, draw=black] (-2,-1.5)circle (2.5pt);
\filldraw[fill=black, draw=black] (-1.5,-1.5)circle (2.5pt);
\filldraw[fill=black, draw=black] (-1,-1.5)circle (2.5pt);
\filldraw[fill=black, draw=black] (-0.5,-1.5)circle (2.5pt);
\filldraw[fill=black, draw=black] (2,-1.5)circle (2.5pt);
\filldraw[fill=black, draw=black] (2.5,-1.5)circle (2.5pt);
\filldraw[fill=black, draw=black] (3,-1.5)circle (2.5pt);
\filldraw[fill=black, draw=black] (0,-2)circle (2.5pt);
\filldraw[fill=black, draw=black] (0.5,-2)circle (2.5pt);
\filldraw[fill=black, draw=black] (1,-2)circle (2.5pt);
\filldraw[fill=black, draw=black] (1.5,-2)circle (2.5pt);
\filldraw[fill=black, draw=black] (-2,-2)circle (2.5pt);
\filldraw[fill=black, draw=black] (-1.5,-2)circle (2.5pt);
\filldraw[fill=black, draw=black] (-1,-2)circle (2.5pt);
\filldraw[fill=black, draw=black] (-0.5,-2)circle (2.5pt);
\filldraw[fill=black, draw=black] (2,-2)circle (2.5pt);
\filldraw[fill=black, draw=black] (2.5,-2)circle (2.5pt);
\filldraw[fill=black, draw=black] (3,-2)circle (2.5pt);
\filldraw[fill=black, draw=black] (0,-2.5)circle (2.5pt);
\filldraw[fill=black, draw=black] (0.5,-2.5)circle (2.5pt);
\filldraw[fill=black, draw=black] (1,-2.5)circle (2.5pt);
\filldraw[fill=black, draw=black] (1.5,-2.5)circle (2.5pt);
\filldraw[fill=black, draw=black] (-2,-2.5)circle (2.5pt);
\filldraw[fill=black, draw=black] (-1.5,-2.5)circle (2.5pt);
\filldraw[fill=black, draw=black] (-1,-2.5)circle (2.5pt);
\filldraw[fill=black, draw=black] (-0.5,-2.5)circle (2.5pt);
\filldraw[fill=black, draw=black] (2,-2.5)circle (2.5pt);
\filldraw[fill=black, draw=black] (2.5,-2.5)circle (2.5pt);
\filldraw[fill=black, draw=black] (3,-2.5)circle (2.5pt);
\filldraw[fill=black, draw=black] (0,-3)circle (2.5pt);
\filldraw[fill=black, draw=black] (0.5,-3)circle (2.5pt);
\filldraw[fill=black, draw=black] (1,-3)circle (2.5pt);
\filldraw[fill=black, draw=black] (1.5,-3)circle (2.5pt);
\filldraw[fill=black, draw=black] (-2,-3)circle (2.5pt);
\filldraw[fill=black, draw=black] (-1.5,-3)circle (2.5pt);
\filldraw[fill=black, draw=black] (-1,-3)circle (2.5pt);
\filldraw[fill=black, draw=black] (-0.5,-3)circle (2.5pt);
\filldraw[fill=black, draw=black] (2,-3)circle (2.5pt);
\filldraw[fill=black, draw=black] (2.5,-3)circle (2.5pt);
\filldraw[fill=black, draw=black] (3,-3)circle (2.5pt);
\filldraw[fill=black, draw=black] (0,-3.5)circle (2.5pt);
\filldraw[fill=black, draw=black] (0.5,-3.5)circle (2.5pt);
\filldraw[fill=black, draw=black] (1,-3.5)circle (2.5pt);
\filldraw[fill=black, draw=black] (1.5,-3.5)circle (2.5pt);
\filldraw[fill=black, draw=black] (-2,-3.5)circle (2.5pt);
\filldraw[fill=black, draw=black] (-1.5,-3.5)circle (2.5pt);
\filldraw[fill=black, draw=black] (-1,-3.5)circle (2.5pt);
\filldraw[fill=black, draw=black] (-0.5,-3.5)circle (2.5pt);
\filldraw[fill=black, draw=black] (2,-3.5)circle (2.5pt);
\filldraw[fill=black, draw=black] (2.5,-3.5)circle (2.5pt);
\filldraw[fill=black, draw=black] (3,-3.5)circle (2.5pt);
\filldraw[fill=black, draw=black] (0,-4)circle (2.5pt);
\filldraw[fill=black, draw=black] (0.5,-4)circle (2.5pt);
\filldraw[fill=black, draw=black] (1,-4)circle (2.5pt);
\filldraw[fill=black, draw=black] (1.5,-4)circle (2.5pt);
\filldraw[fill=black, draw=black] (-2,-4)circle (2.5pt);
\filldraw[fill=black, draw=black] (-1.5,-4)circle (2.5pt);
\filldraw[fill=black, draw=black] (-1,-4)circle (2.5pt);
\filldraw[fill=black, draw=black] (-0.5,-4)circle (2.5pt);
\filldraw[fill=black, draw=black] (2,-4)circle (2.5pt);
\filldraw[fill=black, draw=black] (2.5,-4)circle (2.5pt);
\filldraw[fill=black, draw=black] (3,-4)circle (2.5pt);
\filldraw[fill=red!20, draw=red, thick] (-1.5,0)--(1.5,0)--(1.5,-3)--(-1.5,-3)--cycle;
\filldraw[fill=green!20, draw=black] (-2,0)circle (7pt);
\filldraw[fill=green!20, draw=black] (-1.5,-3.5)circle (7pt);
\filldraw[fill=green!20, draw=black] (1.5,-3.5)circle (7pt);
\filldraw[fill=green!20, draw=black] (-2,-3)circle (7pt);
\path (1.5,-3.5) node [style=sergio] {$A_2$};
\path (-1.5,-3.5) node [style=sergio] {$A_1$};
\path (-2,-3) node [style=sergio] {$A_4$};
\path (-2,0) node [style=sergio] {$A_3$};
\end{tikzpicture}
\caption{Coupled-wire model with 2-foliated subsystem symmetries $U_x(g^y)$ and $U_y(g^x)$. Blue strips depict subsystem symmetries, and $A_{1,2,3,4}$ depict the background gauge fields of corresponding subsystem symmetries marked by red solid lines. 
}
\label{2dSSym}
\end{figure}
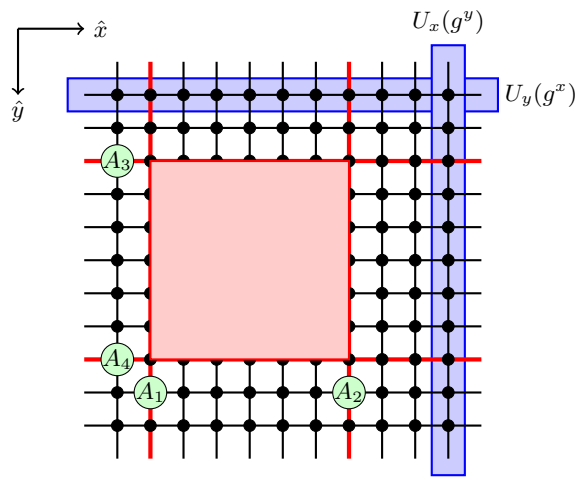

Consider a system with finite extension in $x$ and $y$, and infinite in the remaining directions. The boundary consists of two lines/planes parallel to $x$, and two parallel to $y$, with four (codimension-2) corners. By definition, the boundary is trivially gapped except at the corners. Each corner individually can be described by a theory in $(d-2)$-dimensional space. The subsystem symmetries, when restricted to one corner, become internal global symmetries of the corner theory. For homogeneous subsystem symmetry, the two subsystem symmetries should have identical symmetry action in the corner theory, since the corner is where two submanifolds intersect. Label the four corners as BL, BR, TL, and TR (B for bottom, T for top, L for left, and R for right). The corresponding submanifolds passing through the four corners are labeled as 1, 2, 3, and 4, as illustrated in Fig. \ref{2dSSym}.

We now view this whole system as a $(d-2)$-dimensional system,  with a on-site symmetry group $G_s^{\times 4}$. Clearly, as a physical $(d-2)$-dimensional system, the symmetry action of $G_s^{\times 4}$ must be free of any 't Hooft anomaly. This is the consistency condition that we will impose to classify the SSPT phases. But before going to the actual classification, first we review how to describe 't Hooft anomaly.

For a local quantum system in $(D+1)$-dimensions (may be a lattice model or a continuum theory) with global unitary symmetry group $G$, the 't Hooft anomaly of $G$ can be probed by coupling the system to a (flat) background $G$ gauge field $A$. The anomaly is the fact that the system is not invariant under gauge transformations of the background gauge field $A$. Through the inflow mechanism, the anomaly can be uniquely associated with a $(D+2)$-dimensional invertible theory (the background gauge field should also be extended to the $(D+2)$-dimensional bulk). The topological response theory of the bulk to $A$ will be denoted by $S[A]$, which should be a quantized topological term of the background gauge field $A$. For bosonic systems in $D\leq 2$ with unitary symmetry, the anomaly (and the associated SPT phases) can be fully classified by the group cohomology $\H^{D+2}(G, \U)$. Namely, each anomaly class (or the SPT phase) is uniquely determined by a cohomology class $[\nu]\in \H^{D+2}(G, \U)$, where $\nu$ is a representative group cocycle. Formally, the anomaly action can be written as
\begin{equation}
    S_\text{anomaly}[A]=\int_{M_{D+2}} A^*\nu,
\end{equation}
where we view the background gauge field $A$ as a mapping from the spacetime manifold $M_{D+2}$ to the classifying space $BG$, and $A^*\nu$ is the pullback of $\H^d(BG, \mathbb{R}/\Z)$.

In our case, since the symmetry group is $G_s^{\times 4}$, the background gauge fields can be written as $(A_1, A_2, A_3, A_4)$, where the $1,2,3,4$ indices indicate which subsystem the symmetry acts on, but we should keep in mind that in the $(d-2)$-dimensional system they all become internal global symmetries. Since $G_s$ is Abelian, we adopt the convention that the gauge field $A_i$'s are all additive.

The structure of the system symmetry places more constraints on how the symmetries act on the corner theories. To give one example, consider the BL corner. By definition, $G_s^{(3)}$ and $G_s^{(4)}$ do not act on BL corner (so the theory is not coupled to $A_3$ and $A_4$), and $G_s^{(1)}$ and $G_s^{(2)}$ have identical symmetry actions. So when $A_1$ (or $A_2$) is turned on, the 't Hooft anomaly of the BL theory is captured by an action $S_\text{BL}[A_1]$ ($S_\text{BL}[A_2])$. However, when both $A_1$ and $A_2$ are turned on, the anomaly response of the BL corner theory becomes $S_\text{BL}[A_1+A_2]$. 

Now we consider the anomaly of $G_s^{(1)}$ alone. Equivalently, only $A_1$ is turned on. Both TL and BL corner theories are coupled to $A_1$, so the vanishing of the anomaly for $G_s^{(1)}$ implies
\begin{equation}
    S_\text{BL}[A_1]+S_\text{TL}[A_1]=0.
\end{equation}
Similarly, we have 
\begin{equation}
\begin{split}
    S_\text{BR}[A_2]+S_\text{TR}[A_2]=0,\\
    S_\text{TL}[A_3]+S_\text{TR}[A_3]=0,\\
    S_\text{BL}[A_4]+S_\text{BR}[A_4]=0.
\end{split}
\end{equation}
Thus the four response actions can all be related to e.g. $S_\text{BL}$:
\begin{equation}
    \begin{split}
        S_\text{BR}[A]=-S_\text{BL}[A],\\
        S_\text{TL}[A]=-S_\text{BL}[A],\\
        S_\text{TR}[A]=S_\text{BL}[A].
    \end{split}
\end{equation}
Next we consider $G_s^{(1)}\times G_s^{(4)}$, i.e. turning on both $A_1$ and $A_4$, which intersect at the BL corner. The vanishing of the anomaly leads to 
\begin{equation}
    S_\text{BL}[A_1+A_4]+S_\text{TL}[A_1]+S_\text{BR}[A_4]=0.
\end{equation}
Using the relations, we have
\begin{equation}
     S_\text{BL}[A_1+A_4]-S_\text{BL}[A_1]-S_\text{BL}[A_4]=0.
     \label{noanomaly}
\end{equation}
It is important that the left-hand side should be viewed as the anomaly response for the symmetry group $G_s\times G_s$.

It is easy to see that considering other pairs of intersecting subsystem symmetries lead to the same mathematical condition as \eqref{noanomaly}. Furthermore, once \eqref{noanomaly} is satisfied, the symmetry group $G_s^{\times 4}$ is indeed non-anomalous.

We now translate the condition \eqref{noanomaly} into a more concrete statement about the group cocycle. Suppose the anomaly action $S_\text{BL}$ corresponds to a group cocycle $[\nu]\in \H^{d}(G_s, \U)$.  We introduce an ``obstruction" map from $\H^d(G_s, \U)$ to $\H^d(G_s\times G_s, \U)$, as
\begin{equation}
    f(\nu)((g_1, g_1'), \cdots, (g_d,g_d'))=\frac{\nu(g_1,\cdots,g_d)\nu(g_1',\cdots, g_d')}{\nu(g_1g_1',\cdots, g_dg_d')},
    \label{2-foliated anomaly free1}
\end{equation}
and the condition \eqref{noanomaly} is the statement that $[f(\nu)]$ is trivial in $\H^d(G_s\times G_s, \U)$, which defines a subgroup of $\H^d(G_s, \U)$.

We note that for $d=2$, the corner theory is 0+1d and the only anomaly is the 't Hooft anomaly of a global symmetry, which essentially says the corner state transforms as a projective representation of the global symmetry. In $d=3$, the corner theory is 1+1d, which may have a gravitational anomaly characterized by a chiral central charge $c$, which is always an integer multiple of $8$ in bosonic systems. However, in that case, the gravitational anomaly and 't Hooft anomaly can be completely decoupled, and the vanishing of the gravitational anomaly requires
\begin{equation}
    c_\text{BL}+c_\text{TL}+c_\text{BR}+c_\text{TR}=0.
\end{equation}
Then they can always be cancelled by stacking layers of $E_8$ states on the four surfaces.

Next, we turn to the case with an additional global symmetry $G_g$. The story is similar but with a few twists. We will also need to turn on background gauge fields for the global symmetry, and the corner theory can have mixed anomalies between the global and the subsystem symmetries. We can assume that the corner theories do not carry any anomaly of the $G_g$ symmetry alone, as they can be cancelled by boundary reconstruction. This is analogous to the argument that we can ignore the gravitational anomaly. 

Therefore, the anomaly response for e.g. the BL corner theory in the presence of both subsystem background gauge field $A_s$ and $A_g$ should take the following form:
\begin{equation}
    S_\text{BL}[A_s]+S_\text{BL}[A_s, A_g].
\end{equation}
The first term is the subsystem symmetry anomaly, which we have already studied carefully.
The second term represents the mixed anomaly. Namely, it is only nontrivial when both $A_s$ and $A_g$ are nontrivial.

Following the same anomaly vanishing argument, we find
\begin{equation}
    \begin{split}
        S_\text{BR}[A_s, A_g]=-S_\text{BL}[A_s, A_g]\\
        S_\text{TL}[A_s, A_g]=-S_\text{BL}[A_s, A_g]\\
        S_\text{TR}[A_s, A_g]=S_\text{BL}[A_s, A_g].
    \end{split}
\end{equation}
and
\begin{equation}
     S_\text{BL}[A_s+A_s', A_g]-S_\text{BL}[A_s, A_g]-S_\text{BL}[A_s', A_g]=0.
     \label{nomixedanomaly}
\end{equation}

Let us turn it into an algebraic expression for the case of group-cohomology SPT phases. We assume that $S_\text{BL}[A_s, A_g]$ corresponds to a cocycle $\nu$ in $\H^d(G_s\times G_g, \U)$. Denote the group elements of $G_s\times G_g$ by a pair $(g, h)$ where $g\in G_s, h\in G_g$. Define the obstruction map from $\nu$ to $\H^d(G_s\times G_s\times G_g, \U)$:
\begin{equation}
\begin{split}
    f&(\nu)\left( (g_1, g_1', h_1), \cdots, (g_d, g_d', h_d)\right)=\\
    &\frac{\nu\left( (g_1, h_1), \cdots, (g_d, h_d)\right)\nu\left( (g_1', h_1), \cdots, (g_d', h_d)\right)}{\nu\left( (g_1g_1', h_1), \cdots, (g_dg_d', h_d)\right)}.
\end{split}
\label{2-foliated anomaly free2}
\end{equation}
Here we denote the elements of $G_s\times G_s\times G_g$ as $(g, g', h)$.

The condition is then $f(\nu)$ corresponds to a trivial class in $\H^d(G_s\times G_s\times G_g, \U)$.

\subsection{Microscopic constructions}
Next, we demonstrate that all obstruction-vanishing anomalies can be realized in the corner theory of a bulk SSPT state, generalizing a ``coupled wire" construction in \cite{ teo2014luttinger}.

\begin{figure}
\begin{tikzpicture}[scale=0.9]
\tikzstyle{sergio}=[rectangle,draw=none]
\filldraw[fill=green!20, draw=green, thick] (-0.5,0.5)--(2,0.5)--(2,-2)--(-0.5,-2)--cycle;
\filldraw[fill=green!20, draw=none, thick] (-0.5,2.5)--(2,2.5)--(2,1.5)--(-0.5,1.5)--cycle;
\draw[draw=green,thick] (-0.5,2.5) -- (-0.5,1.5);
\draw[draw=green,thick] (2,1.5) -- (-0.5,1.5);
\draw[draw=green,thick] (2,1.5) -- (2,2.5);
\filldraw[fill=green!20, draw=none, thick] (4,-2)--(4,0.5)--(3,0.5)--(3,-2)--cycle;
\draw[draw=green,thick] (4,-2) -- (3,-2);
\draw[draw=green,thick] (3,-2) -- (3,0.5);
\draw[draw=green,thick] (3,0.5) -- (4,0.5);
\filldraw[fill=green!20, draw=none, thick] (4,1.5)--(3,1.5)--(3,2.5)--(4,2.5)--cycle;
\draw[draw=green,thick] (3,2.5) -- (3,1.5);
\draw[draw=green,thick] (4,1.5) -- (3,1.5);
\filldraw[fill=yellow!20, draw=yellow,thick] (-1.5,1) ellipse (12pt and 26pt);
\filldraw[fill=yellow!20, draw=yellow,thick] (-1.5,-2.5) ellipse (12pt and 26pt);
\filldraw[fill=yellow!20, draw=yellow,thick] (2.5,-3) ellipse (26pt and 12pt);
\draw[thick] (-2.5,1) -- (4,1);
\draw[thick] (2.5,-4) -- (2.5,2.5);
\draw[thick] (-1,2.5) -- (-1,-4);
\draw[thick] (-2.5,-2.5) -- (4,-2.5);
\draw[thick] (-2,2) -- (0,2);
\draw[thick] (-2,2) -- (-2,0);
\draw[thick] (0,0) -- (0,2);
\draw[thick] (0,0) -- (-2,0);
\draw[thick] (3.5,2) -- (1.5,2);
\draw[thick] (1.5,0) -- (1.5,2);
\draw[thick] (3.5,2) -- (3.5,0);
\draw[thick] (3.5,0) -- (1.5,0);
\draw[thick] (0,-1.5) -- (-2,-1.5);
\draw[thick] (-2,-3.5) -- (-2,-1.5);
\draw[thick] (0,-1.5) -- (0,-3.5);
\draw[thick] (0,-3.5) -- (-2,-3.5);
\draw[thick] (1.5,-1.5) -- (3.5,-1.5);
\draw[thick] (1.5,-1.5) -- (1.5,-3.5);
\draw[thick] (3.5,-3.5) -- (3.5,-1.5);
\draw[thick] (1.5,-3.5) -- (3.5,-3.5);
\path (-1.5,0.5) node [style=sergio] {\large $\bigotimes$};
\path (-0.5,0.5) node [style=sergio] {\large $\bigodot$};
\path (-0.5,1.5) node [style=sergio] {\large $\bigotimes$};
\path (-1.5,1.5) node [style=sergio] {\large $\bigodot$};
\path (2,0.5) node [style=sergio] {\large $\bigotimes$};
\path (3,0.5) node [style=sergio] {\large $\bigodot$};
\path (3,1.5) node [style=sergio] {\large $\bigotimes$};
\path (2,1.5) node [style=sergio] {\large $\bigodot$};
\path (-1.5,-3) node [style=sergio] {\large $\bigotimes$};
\path (-0.5,-3) node [style=sergio] {\large $\bigodot$};
\path (-0.5,-2) node [style=sergio] {\large $\bigotimes$};
\path (-1.5,-2) node [style=sergio] {\large $\bigodot$};
\path (2,-3) node [style=sergio] {\large $\bigotimes$};
\path (3,-3) node [style=sergio] {\large $\bigodot$};
\path (3,-2) node [style=sergio] {\large $\bigotimes$};
\path (2,-2) node [style=sergio] {\large $\bigodot$};
\path (-1.25,2.75) node [style=sergio] {$A_1^v$};
\path (4.4,-2.5) node [style=sergio] {$A_n^h$};
\path (4.5,1) node [style=sergio] {$A_{n-1}^h$};
\path (2.75,2.75) node [style=sergio] {$A_2^v$};
\draw[thick] (-3,2.5) -- (-3,-3);
\filldraw[fill=blue!20, draw=blue, thick] (-3.3,-1.65)--(-2.7,-1.65)--(-2.7,-3.35)--(-3.3,-3.35)--cycle;
\filldraw[fill=blue!20, draw=blue, thick] (-3.3,1.85)--(-2.7,1.85)--(-2.7,0.15)--(-3.3,0.15)--cycle;
\filldraw[fill=red!20, draw=red,thick] (-3,-0.75) ellipse (15pt and 48pt);
\path (-3,-3) node [style=sergio] {\large $\bigotimes$};
\path (-3,-2) node [style=sergio] {\large $\bigodot$};
\path (-3,0.5) node [style=sergio] {\large $\bigotimes$};
\path (-3,1.5) node [style=sergio] {\large $\bigodot$};
\end{tikzpicture}
\caption{Illustration of the ``coupled wire" construction. Each ``$\bigotimes$'' represents a codimension-2 system with the given anomaly of $G_s$ (e.g. a projective representation in $d=2$, or a Luttinger liquid in $d=3$), and each ``$\bigodot$'' represents a system with the opposite anomaly. Each square represents a ``site", which has no anomaly for the $G_s$ symmetry and thus can be realized by certain microscopic model. Green plaquettes represent the four-body interactions in bulk, and yellow ellipse for the on-site two-body coupling on the edge. $(A_1^v,A_2^v,A_{n-1}^h,A_n^h)$ are background gauge fields of subsystem symmetry $G_s$ defined on different codimension-1 subsystems. On the left edge, a codimension-1 SSPT phase might be adhered to trivialize the bulk SSPT phase, each blue plate represents a site and each red ellipse represents a two-body inter-site coupling.}
\label{hinge}
\end{figure}
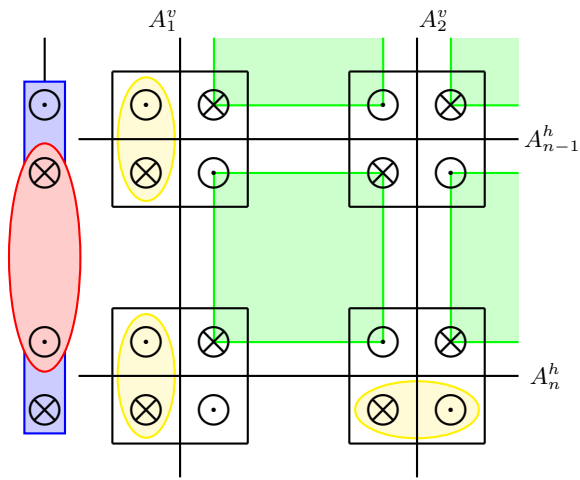


For this, we can consider the following lattice construction. First we pick a $(d-2)$-dimensional symmetry-preserving gapless theory $\mathcal{T}$ with the given $G_s$ anomaly class $S[A]$. For $d=2$, it is a projective representation.  For $d=3$, one can choose a (1+1)d CFT with the given anomaly. Denote by $\overline{\mathcal{T}}$ the theory with the opposite anomaly (e.g. complex conjugating $\mathcal{T}$). Given a square grid as shown in Fig. \ref{hinge}, in each unit cell we arrange a $(d-2)$-dimensional theory $\mathcal{T}_\text{BL}\otimes \overline{\mathcal{T}}_\text{TL}\otimes \overline{\mathcal{T}}_\text{BR}\otimes \mathcal{T}_\text{TR}$. Within each unit cell, the symmetry $G_s$ acts diagonally on the four theories. By construction, the unit cell has no $G_s$ anomaly, so it should be possible to realize the theory in a physical $(d-2)$ system with on-site $G_s$ symmetry. 

Now we consider the whole grid. The subsystem symmetry is defined in the standard way. To construct the SSPT state, we turn on interactions at each square plaquette, coupling the four neighboring unit cells indicated by the green area in Fig. \ref{hinge}. Note that the coupling only involves one of the $\mathcal{T}$ or $\overline{\mathcal{T}}$ theory from each of the four unit cells. We require that the four theories involved in the plaquette interaction can be gapped out while preserving the subsystem symmetries. In order to do this, the subsystem symmetries acting on this green plaquette must be non-anomalous, which is precisely the condition previously in \eqref{noanomaly}. In fact, it is believed that anomaly vanishing is both sufficient and necessary in order to find a trivially gapped ground state\cite{JWXGWanomaly2013}. So there should exist such an interaction. 
 
Once the appropriate interactions are turned on and drive the bulk into a fully gapped state, there still exist nontrivial boundary modes. We assume that the system is terminated with ``smooth" boundaries, as illustrated in Fig. \ref{hinge}. For each unit cell on the smooth edge, there are two dangling theories that are not included in the bulk plaquette interactions, with opposite anomalies (see Fig. \ref{hinge}). Hence we can simply introduce a coupling (see yellow ellipse in Fig. \ref{hinge}) between them preserving the subsystem symmetries to gap them out and obtain a fully-gapped boundary. However, for the unit cell at the corner, there are three dangling theories that are not included in the bulk plaquette interactions, and two of them can be gapped by the above boundary couplings, leaving a dangling corner/hinge theory that carries the desired anomaly. Therefore, this lattice construction gives a symmetric gapped bulk state with symmetric gapped edges but anomalous corner modes. 

\section{Examples\label{2-foliated example}}

\subsection{2D systems with only subsystem symmetry\label{proposition}}
In this subsection, we prove:

\begin{framed}
\textit{There is no nontrivial 2-foliated higher-order SSPT phase in (2+1)D without global symmetry.}
\end{framed}

Let us think about abelian subsystem symmetry. It is well-known that any finite Abelian group can be written as a product of several cyclic groups: 
\begin{align}
G_s=\prod\limits_{i=1}^N\mathbb{Z}_{n_i},~n_i\in\mathbb{Z}
\end{align}
A group element $a\in G_s$ can be expressed as:
\begin{align}
a=\left(a_1,a_2,\cdot\cdot\cdot,a_N\right),~a_{n_i}\in\mathbb{Z}_{n_i}
\end{align}
The general expression of 2-cocycles in $\mathcal{H}^2[G_s,U(1)]$ is:
\begin{align}
\nu_2(a,b)=\exp\left\{2\pi i\sum\limits_{i<j}\frac{p_{ij}}{n_{ij}}a_ib_j\right\}
\label{2-cocycles}
\end{align}
where $p_{ij}\in\mathbb{Z}$ and $n_{ij}\in\mathbb{Z}$ is the greatest common divisor (GCD) of $n_i$ and $n_j$. Substitute this explicit 2-cocycle into the obstruction-free condition (\ref{noanomaly}), we conclude that an obstruction-free 2-cocycle corresponding to a nontrivial (2+1)D higher-order SSPT phase requires the following expression to be a 2-coboundary of the $G_s^{2}$ group:
\begin{align}
\nu_2'[(a,a'),(b,b')]=\exp\left\{2\pi i\sum\limits_{i<j}\frac{p_{ij}}{n_{ij}}\left(a_ib_j'+a_i'b_j\right)\right\}
\label{(2+1)D obstructed}
\end{align}
where a group element of $G_s^2$ is expressed as $(a, a')$, $a, a'\in G_s$. To determine if the 2-cocycle (\ref{(2+1)D obstructed}) is a nontrivial 2-cocycle in $\mathcal{H}^2[G_s^2, U(1)]$, we should justify that $\nu_2'$ is not commute for some $(a,a')$ and $(b,b')\in G_s^2$ as
\begin{align}
\nu_2'[(a,a'),(b,b')]\ne\nu_2'[(b,b'),(a,a')]
\end{align}
With the explicit form of 2-cocycle in Eq. (\ref{(2+1)D obstructed}), we have
\[
\nu_2'[(b,b'),(a,a')]=\exp\left\{2\pi i\sum\limits_{i<j}\frac{p_{ij}}{n_{ij}}\left(b_ia_j'+b_i'a_j\right)\right\}
\]
which is not equal to Eq. (\ref{(2+1)D obstructed}) if not all $n_{ij}=1$ for $\forall i,j$. As a consequence, we have proved that all 2-cocycles like Eq. (\ref{2-cocycles}) are obstructed, and there is no nontrivial 2-foliated higher-order SSPT phase in (2+1)D systems without global symmetry, for all Abelian subsystem symmetries.

Furthermore, we have argued that for non-Abelian subsystem symmetry, it is sufficient to consider its Abelian subgroup $G_s'=G_s/[G_s, G_s]$, hence even for non-Abelian subsystem symmetry, there is no (2+1)D higher-order SSPT phase without the aids of some proper global symmetry.

\subsection{2D bosonic SSPT with both subsystem and global symmetries}

The mixed anomaly between $G_s$ and $G_g$ in this case is classified by
    $\omega\in\H^1[G_s, \H^1[G_g, \U]]$.
Let us write the 2-cocycle down explicitly. $\omega$ can be viewed as a group homomorphism between $G_s$ and $\H^1(G_g, \U)$, the latter is the group of one-dimensional representations on $G_g$. Thus the corresponding 2-cocycle can be written as
\begin{align}
\nu( (g_s,g_g), (h_s, h_g))=[\omega(g_s)](h_g).
\label{(2+1)D cup}
\end{align}
Here $\omega(g_s)$ gives a one-dimensional representation of $G_g$ and evaluating it on $h_g$ yields the 2-cocycle. The obstruction is given by
\begin{widetext}
\begin{align}
f(\nu)( (g_s, g_s', g_g), (h_s, h_s', h_g))&=\frac{\nu((g_s, g_g), (h_s, h_g))\nu((g_s',g_g), (h_s', h_g))}{\nu((g_s g_s',g_g), (h_sh_s',h_g))}= \frac{[\omega(g_s)](h_g)[\omega(g_s')](h_g)}{[\omega(g_s g_s')](h_g)}\nonumber\\
&=\left[\frac{\omega(g_s)\omega(g_s')}{\omega(g_sg_s')}\right](h_g)=1.
\end{align}
\end{widetext}
Thus the obstruction vanishes automatically.

Here we present an example to make our construction and classification of SSPT phases more concrete. $G_s=\Z_2$ and  $G_g=\mathbb{Z}_2$. There is a two-dimensional projective representation protected by both $G_s$ and $G_g$. We will construct a 2D SSPT phase with the projective representation as the corner mode.

We use the lattice construction in Fig. \ref{hinge}, where each circle is a spin-$1/2$ degree of freedom. The subsystem symmetry is generated by products of $\sigma_{\bs{R}}^z$ along rows and columns, and the global $\Z_2$ symmetry is generated by $\prod_{\bs{R}}\sigma^x_{\bs{R}}$.
The green plate in Fig. \ref{hinge} now includes four spin-1/2's, and we can introduce a ring-exchange coupling in each plaquette:
\begin{align}
H_r=-\sum\limits_{\bs{R}}\sigma_{\bs{R}}^+\sigma_{\bs{R}+\hat{x}}^-\sigma_{\bs{R}+\hat{x}+\hat{y}}^+\sigma_{\bs{R}+\hat{y}}^-+\text{h.c}.
\end{align}
where $\sigma^{\pm}=\sigma^x+i\sigma^y$, $\bs{R}$ labels the lattice sites and $\hat{x}/\hat{y}$ represents the unit vector along the $x/y$-direction. 

This Hamiltonian has the unique ground state on the lattice with periodic boundary condition (PBC) \cite{YZYou2018}:
\begin{align}
|\Psi_0\rangle=\prod\limits_{P}&\left(|\downarrow_{\mathrm{BL}},\uparrow_{\mathrm{TL}},\uparrow_{\mathrm{BR}},\downarrow_{\mathrm{TR}}\rangle_P\right.\nonumber\\
&\left.+|\uparrow_{\mathrm{BL}},\downarrow_{\mathrm{TL}},\downarrow_{\mathrm{BR}},\uparrow_{\mathrm{TR}}\rangle_P\right)
\end{align}
where the subscript $P$ depicts different plaquettes. For sites on the edge and corner, we turn on a Heisenberg interaction $\bs{\sigma}\cdot\bs{\sigma}'$ in each yellow ellipse in Fig. \ref{hinge},  which can gap out a pair of spin-1/2 degrees of freedom on all edge sites, except one at the corner. As the consequence, there is only one dangling spin-1/2 degree of freedom at each corner of the lattice that remains gapless. 

We notice that the Hamiltonian in fact has more symmetries. For example, the subsystem symmetry group can be enlarged to $\U$, generated by the total $\sigma^z$ along each row and column.  


\subsection{3D bosonic SSPT with $G_s=\Z_2$}

Let us work out the classification of 2nd-order SSPT phases in 3D for $G_s=\Z_2$ and no global symmetry. Since $\H^3(\Z_2, \U)=\Z_2$, we just need to check whether the obstruction vanishes for the nontrivial class. Denote $\Z_2=\{1, g\}$, and the nontrivial class is
\begin{equation}
    \nu(g,g,g)=-1.
\end{equation}
The obstruction mapping  gives a $\H^3$ class $f(\nu)$ for the group $\Z_2\times\Z_2$. We will check the three invariants of the cohomology class:
\begin{equation}
\begin{split}
    f(\nu)&((g,1),(g,1),(g,1))=1\\
    f(\nu)&((1,g),(1,g),(1,g))=1,\\
    f(\nu)&((g,g),(g,g),(g,g))=\nu(g,g,g)^2=1.
\end{split}
\end{equation}
Therefore $f(\nu)$ belongs to the trivial class and the obstruction vanishes. We conclude that the classification is given by $\Z_2$.

Below we provide an explicit construction of the SSPT phase using a coupled wire model.  The (1+1)D system as the building block of the coupled-wire model can only be the edge theory of the (2+1)D Levin-Gu model \cite{LevinGu}, which can be represented in terms of a 2-component Luttinger theory:
\begin{align}
\mathcal{L}_0=\frac{1}{2\pi}\partial_x\phi_1 \partial_\tau\phi_2 + \frac{1}{4\pi}\sum_{\alpha,\beta=1,2}\partial_x\phi_\alpha V_{\alpha\beta}\partial_x\phi_\beta.
\label{Levin-Gu}
\end{align}
 The $\mathbb{Z}_2$ symmetry is generated by the following action:
\begin{align}
\phi_1\rightarrow\phi_1+\pi, \phi_2\rightarrow \phi_2+\pi.
\end{align}

Consider the green plaquette in Fig. \ref{hinge} including four Luttinger liquids (\ref{Levin-Gu}), the overall Lagrangian of these four Luttinger liquids is:
\begin{align}
\mathcal{L}=\frac{1}{4\pi}\partial_x\Phi^{\mathrm{T}}K\partial_\tau\Phi+\frac{1}{4\pi}\partial_x\Phi^{\mathrm{T}}V\partial_x\Phi.
\label{4 copies}
\end{align}
where $\Phi=\left(\phi_1,\cdots,\phi_8\right)^\mathrm{T}$ is the 8-component boson field, $K=\left(\sigma^x\right)^{\oplus4}$ is the $K$-matrix. There are four $\mathbb{Z}_2$ subsystem symmetries $\mathbb{Z}_2^j$ ($j=1,2,3,4$) defined in different directions. Their actions on the bosonic fields, following from, all take the form $\Phi\rightarrow \Phi+\delta\Phi$, where
\begin{align}
\begin{aligned}
&\delta\Phi^{\mathbb{Z}_2^1}=\pi(1,1,1,1,0,0,0,0)^{\mathrm{T}}\\
&\delta\Phi^{\mathbb{Z}_2^2}=\pi(0,0,0,0,1,1,1,1)^{\mathrm{T}}\\
&\delta\Phi^{\mathbb{Z}_2^3}=\pi(1,1,0,0,0,0,1,1)^{\mathrm{T}} \\ 
&\delta\Phi^{\mathbb{Z}_2^4}=\pi(0,0,1,1,1,1,0,0)^{\mathrm{T}}
\end{aligned}
\label{Z2dPhi}
\end{align}
We then need to construct gapping terms that gap out the edge without breaking symmetry, neither explicitly nor spontaneously. Consider backscattering terms of the form:
\begin{align}
U=U_0\sum\limits_{k}\cos\left(l_k^\mathrm{T}K\Phi\right).
\label{Higgs}
\end{align}
Since there are 8 bosonic fields, four independent, mutually commuting gapping terms are needed to completely gap out the edge. More precisely, the vectors $\{l_k\}$ must satisfy the ``null-vector'' conditions \cite{Haldane1995} for $\forall i,j$:
\begin{align}
l_i^\mathrm{T}Kl_j=0.
\label{null-vector}
\end{align}
In addition, the interactions must preserve the $\Z_2$ symmetries, which means for each $i$
\begin{equation}
    l_i^\mathrm{T}K\delta\Phi=0,
\end{equation}
for each of the $\delta\Phi$ in Eq. \eqref{Z2dPhi}.

We find the following vectors  satisfy all the requirements:
\begin{align}
\begin{aligned}
&l_1^\mathrm{T}=(1,0,1,0,1,0,1,0)\\
&l_2^\mathrm{T}=(0,1,0,-1,0,-1,0,1)\\
&l_3^\mathrm{T}=(1,0,0,-1,1,0,0,1)\\
&l_4^\mathrm{T}=(0,1,-1,0,0,-1,-1,0).
\end{aligned}
\label{plaquette Higgs}
\end{align}
 Furthermore, in order to obtain a fully-gapped bulk state, we should avoid spontaneous symmetry breaking, which would lead to ground-state degeneracy in each plaquette. Following the method described in \cite{Lu12,Ning21a}, we confirm that the gapping terms leave a unique ground state, thus no spontaneous symmetry breaking. To summarize, we find that the Higgs terms (\ref{plaquette Higgs}) provide a fully-gapped bulk state. We note that all these Higgs terms are four-body interactions, hence the fully gapped bulk state does not have any layered structure.

For the sites on the (2+1)D surface, there are two Luttinger liquids (\ref{Levin-Gu}) that are not included in the bulk interactions (see yellow ellipses in Fig. \ref{hinge}). The total Lagrangian of these two Luttinger liquids is:
\begin{align}
\mathcal{L}_s=\left(\partial_x\phi_s^{\mathrm{T}}\right)\frac{K^s}{4\pi}\left(\partial_\tau\phi_s\right)+\left(\partial_x\phi_s^{\mathrm{T}}\right)\frac{V^s}{4\pi}\left(\partial_x\phi_s\right)
\label{edge Lagrangian}
\end{align}
where $\phi_s^{\mathrm{T}}=(\phi_1,\phi_2,\phi_3,\phi_4)$ is the 4-component chiral boson field, $K^s=(\sigma^x)^{\oplus2}$ is the $K$-matrix. The on-site $\mathbb{Z}_2$ symmetry is defined as:
\begin{align}
W^{\mathbb{Z}_2}=\mathbbm{1}_{4\times4},~\delta\phi=\pi(1,1,1,1)^{\mathrm{T}}
\end{align}
We can simply gap out these two Luttinger liquids by two on-site Higgs terms (\ref{Higgs}) with the following null-vectors:
\begin{align}
\begin{aligned}
&l_1^{\mathrm{T}}=(1,0,1,0)\\
&l_2^{\mathrm{T}}=(0,1,0,-1)
\end{aligned}
\label{surface Higgs}
\end{align}
And a fully-gapped (2+1)D surface state is obtained.

For the sites at the hinge of the system, there are three dangling (1+1)D Luttinger liquids that are not included in the bulk interactions, where two of them can be gapped by the Higgs term like Eq. (\ref{surface Higgs}). Hence the remaining gapless Luttinger liquid with the Lagrangian (\ref{Levin-Gu}) will be the hinge mode of the constructed coupled-wire model.

We should further investigate the stability of the hinge mode in order to make sure that the coupled-wire model we have constructed characterizes a nontrivial (3+1)D SSPT phase. Consider a potential (2+1)D SSPT phase protected by $G_s$ adhering at the left surface of the system (see Fig. \ref{hinge}) as an assembly of (1+1)D Luttinger liquids, each site includes two Luttinger liquids (see blue rectangles in Fig. \ref{hinge}), each of them has the Lagrangian (\ref{Levin-Gu}). On the one hand, they can be gapped on the on-site stage, by introducing the Higgs terms corresponding to the null-vectors in Eq. (\ref{surface Higgs}). On the other hand, we prove that the inter-site coupling of the Luttinger liquids is prohibited by horizontal subsystem symmetry: consider two Luttinger liquids included in a red ellipse in Fig. \ref{hinge}, there are two subsystem symmetries $\mathbb{Z}_2^1$ and $\mathbb{Z}_2^2$ defined on different vertical coordinates along the horizontal direction:
\begin{align}
\begin{aligned}
&\delta\phi^{\mathbb{Z}_2^1}=\pi(1,1,0,0)^{\mathrm{T}}\\
&\delta\phi^{\mathbb{Z}_2^2}=\pi(0,0,1,1)^{\mathrm{T}}
\end{aligned}
\end{align}
We can rigorously prove that these two Luttinger liquids in a red ellipse cannot be gapped \cite{levin18}. Thus the (1+1)D hinge mode as the edge theory of the (2+1)D Levin-Gu state is robust against adhering a (2+1)D SSPT phase to the edge of the system, and we have constructed a nontrivial 2-foliated $\mathbb{Z}_2$ SSPT phase.

\subsection{3D bosonic SSPT with $G_s=\mathbb{Z}_2$ and $G_g=\mathbb{Z}_2^T$}
In this section, we explicit construct the coupled-wire model of (3+1)D second-order SSPT with 2-foliated subsystem symmetry $\mathbb{Z}_2$ and a global time-reversal symmetry $\mathbb{Z}_2^T$. The Luttinger liquid we work on as the building block carries the mixed anomaly of $\mathbb{Z}_2$ and $\mathbb{Z}_2^T$, which is classified by $\mathcal{H}^1\left(\mathbb{Z}_2,\mathcal{H}^2\left[\mathbb{Z}_2^T, U_T(1)\right]\right)$. The two-component Luttinger liquid takes the form of Eq. (\ref{Levin-Gu}), with the following symmetry properties
\begin{align}
\begin{aligned}
&\mathbb{Z}_2:~\phi_1\mapsto\phi_1+\pi,~\phi_2\mapsto\phi_2\\
&\mathbb{Z}_2^T:~\phi_1\mapsto\phi_1,~\phi_2\mapsto-\phi_2+\pi
\end{aligned}
\end{align}
Again consider the green block in Fig. \ref{hinge} including four Luttinger liquids (\ref{Levin-Gu}), the overall Lagrangian takes the form of (\ref{4 copies}). There are four $\mathbb{Z}_2$ subsystem symmetries $\mathbb{Z}_2^j$ ($j=1,2,3,4$) defined in different directions. These $\mathbb{Z}_2$ actions take the form of $\Phi\mapsto\Phi+\delta\Phi$, where 
\begin{align}
\begin{aligned}
&\delta\Phi^{\mathbb{Z}_2^1}=\pi(1,0,1,0,0,0,0,0)^{\mathrm{T}}\\
&\delta\Phi^{\mathbb{Z}_2^2}=\pi(0,0,0,0,1,0,1,0)^{\mathrm{T}}\\
&\delta\Phi^{\mathbb{Z}_2^3}=\pi(1,0,0,0,0,0,1,0)^{\mathrm{T}} \\ 
&\delta\Phi^{\mathbb{Z}_2^4}=\pi(0,0,1,0,1,0,0,0)^{\mathrm{T}}
\end{aligned}
\end{align}
and the global time-reversal symmetry acts as $\Phi\mapsto W^{\mathcal{T}}\Phi+\delta\Phi^{\mathcal{T}}$, where $W^{\mathcal{T}}=(\sigma^z)^{\oplus4}$ and $\delta\Phi^{\mathcal{T}}=\pi(0,1,0,1,0,1,0,1)^{\mathrm{T}}$. In order to gap out all eight components of the boson fields, we need at least four independent Higgs terms (\ref{Higgs}) with four null-vectors $\{l_k\}$. We find that the following null-vectors satisfy all conditions:
\begin{align}
\begin{aligned}
&l_1^{\mathrm{T}}=(1,0,0,0,1,0,0,0)\\
&l_2^{\mathrm{T}}=(1,0,1,0,0,0,0,0)\\
&l_3^{\mathrm{T}}=(0,0,0,0,1,0,1,0)\\
&l_4^{\mathrm{T}}=(0,1,0,1,0,1,0,1)\\
\end{aligned}
\end{align}
which avoids spontaneous symmetry breaking. 

For the sites on the (2+1)D surface, there are two Luttinger liquids (\ref{Levin-Gu}) that are not included in the bulk interactions (see yellow ellipses in Fig. \ref{hinge}). The Lagrangian of these two Luttinger liquids takes the form of Eq. (\ref{edge Lagrangian}). The on-site $\mathbb{Z}_2\times\mathbb{Z}_2^T$ symmetry is defined as
\begin{align}
\begin{aligned}
&W^{\mathbb{Z}_2}=\mathbbm{1}_{4\times4},~\delta\phi^{\mathbb{Z}_2}=\pi(1,0,1,0)^{\mathrm{T}}\\
&W^{\mathbb{Z}_2^T}=(\sigma^z)^{\oplus2},~\delta\phi^{\mathbb{Z}_2}=\pi(0,1,0,1)^{\mathrm{T}}
\end{aligned}
\end{align}
By introducing the following two Higgs terms, we can simply gap out these Luttinger liquids and obtain a fully-gapped (2+1)D surface state
\begin{align}
\begin{aligned}
&l_1^{\mathrm{T}}=(1,0,1,0)\\
&l_2^{\mathrm{T}}=(0,1,0,1)
\end{aligned}
\end{align}

For the sites at the hinge which include three dangling Luttinger liquids (\ref{Levin-Gu}), the above two of them can simply be gapped out by surface Higgs terms. Therefore, the remaining gapless Luttinger liquid should be the second-order hinge mode of the (3+1)D SSPT phase with 2-foliated $\mathbb{Z}_2$ subsystem symmetry and a global time-reversal symmetry $\mathbb{Z}_2^T$ who carries their mixed anomaly.

\subsection{3D fermionic SSPT with $G_s=\Z_2^f$ and $G_g=\mathbb{Z}_2 $}
Our arguments can also be generalized to interacting fermionic systems. We demonstrate this generalization by an example with $G_s=\mathbb{Z}_2^f$ and $G_g=\mathbb{Z}_2$. The (1+1)D Luttinger liquid as the building block of the coupled-wire model is
\begin{align}
\mathcal{L}_0=\frac{1}{4\pi}\partial_x\phi^{\mathrm{T}}\sigma^z\partial_t\phi+\frac{1}{4\pi}\sum\limits_{\alpha,\beta=1,2}\partial_x\phi_{\alpha}V_{\alpha\beta}\partial_x\phi_\beta
\label{fermionic Levin-Gu}
\end{align}
The $\mathbb{Z}_2$ symmetry is defined as
\begin{align}
\phi_1\mapsto\phi_1,~\phi_2\mapsto-\phi_2
\end{align}
and $\mathbb{Z}_2^f$ fermion parity as
\begin{align}
\phi_1\mapsto\phi_1+\pi,~\phi_2\mapsto\phi_2+\pi
\end{align}
Repeatedly consider the green plaquette in Fig. \ref{hinge} composed of four Luttinger liquids (\ref{fermionic Levin-Gu}), forming the overall Lagrangian as
\begin{align}
\mathcal{L}=\frac{1}{4\pi}\partial_x\Phi^{\mathrm{T}}K\partial_\tau\Phi+\frac{1}{4\pi}\partial_x\Phi^{\mathrm{T}}V\partial_x\Phi
\end{align}
where $\Phi=(\phi_1,\cdot\cdot\cdot,\phi_8)^{\mathrm{T}}$ is the 8-component boson field, $K=(\sigma_z)^{\oplus4}$ is the $K$-matrix. There are four $\mathbb{Z}_2^f$ subsystem symmetries $\mathbb{Z}_{2,j}^f$ ($j=1,2,3,4$) in different directions with an additional $\mathbb{Z}_2$ global symmetry. The fermion parities act on the boson fields have the form $\Phi\mapsto\Phi+\delta\Phi$, where
\begin{align}
\begin{aligned}
&\delta\Phi^{\mathbb{Z}_{2,1}^f}=\pi(1,1,1,1,0,0,0,0)^{\mathrm{T}}\\
&\delta\Phi^{\mathbb{Z}_{2,2}^f}=\pi(0,0,0,0,1,1,1,1)^{\mathrm{T}}\\
&\delta\Phi^{\mathbb{Z}_{2,3}^f}=\pi(1,1,0,0,0,0,1,1)^{\mathrm{T}}\\
&\delta\Phi^{\mathbb{Z}_{2,4}^f}=\pi(0,0,1,1,1,1,0,0)^{\mathrm{T}}
\end{aligned}
\end{align}
and the global $\mathbb{Z}_2$ symmetry action takes the form $\Phi\mapsto W\Phi$, where $W=(\sigma^z)^{\oplus4}$. In order to gap out all these boson fields, we need to find four independent Higgs terms like Eq. (\ref{Higgs}) while vectors $\{l_k\}$ ($k=1,2,3,4$) must be null-vectors \cite{Haldane1995}. We find the following null-vectors:
\begin{align}
\begin{aligned}
&l_1^{\mathrm{T}}=(1,0,0,1,1,0,0,1)\\
&l_2^{\mathrm{T}}=(0,1,1,0,0,1,1,0)\\
&l_3^{\mathrm{T}}=(1,1,1,1,0,0,0,0)\\
&l_4^{\mathrm{T}}=(1,1,0,0,1,1,0,0)
\end{aligned}
\end{align}
and the corresponding gapping terms avoid spontaneous symmetry breaking. 

For the sites on the (2+1)D surface, there are two Luttinger liquids (\ref{fermionic Levin-Gu}), as illustrated by yellow ellipses in Fig. \ref{hinge}. These two Luttinger liquids form the following Lagrangian
\begin{align}
\mathcal{L}_s=\partial_x\phi_s^{\mathrm{T}}\frac{K^s}{4\pi}\partial_t\phi_s+\partial_x\phi_s^{\mathrm{T}}\frac{V^s}{4\pi}\partial_x\phi_s
\end{align}
where $\phi_s^{\mathrm{T}}=(\phi_1,\cdot\cdot\cdot,\phi_4)$ is the 4-component boson field and $K^s=(\sigma^z)^{\oplus2}$ is the $K$-matrix. The on-site $\mathbb{Z}_2\times\mathbb{Z}_2^f$ is defined as
\begin{align}
\begin{aligned}
&W^{\mathbb{Z}_2}=(\sigma^z)^{\oplus2},~\delta\phi^{\mathbb{Z}_2}=0\\
&W^{\mathbb{Z}_2^f}=\mathbbm{1}_{4\times4},~\delta\phi^{\mathbb{Z}_2}=\pi(1,1,1,1)
\end{aligned}
\end{align}
These boson fields can simply be gapped out through two Higgs terms (\ref{Higgs}) with the following null-vectors to obtain a fully gapped (2+1)D surface state:
\begin{align}
\begin{aligned}
&l_1^{\mathrm{T}}=(1,0,0,1)\\
&l_2^{\mathrm{T}}=(0,1,1,0)
\end{aligned}
\end{align}

Finally, for the sites at the hinge of the system, there are three dangling (1+1)D Luttinger liquids (\ref{fermionic Levin-Gu}), where two of them can be gapped by the surface Higgs terms. Therefore, the remaining gapless Luttinger liquid will be the hinge mode of the constructed coupled wire model of (3+1)D SSPT phase with 2-foliated $\mathbb{Z}_2^f$ fermion parities and a global $\mathbb{Z}_2$ symmetry. 

\section{Third-order SSPT phases in 3D systems\label{Sec.corner}}
Next, we consider the 3d systems with 3-foliated subsystem symmetries and demonstrate the resulting third-order SSPT phases. 

\subsection{Classification using boundary anomaly}

\begin{figure}
\centering\includegraphics[width=0.48\textwidth]{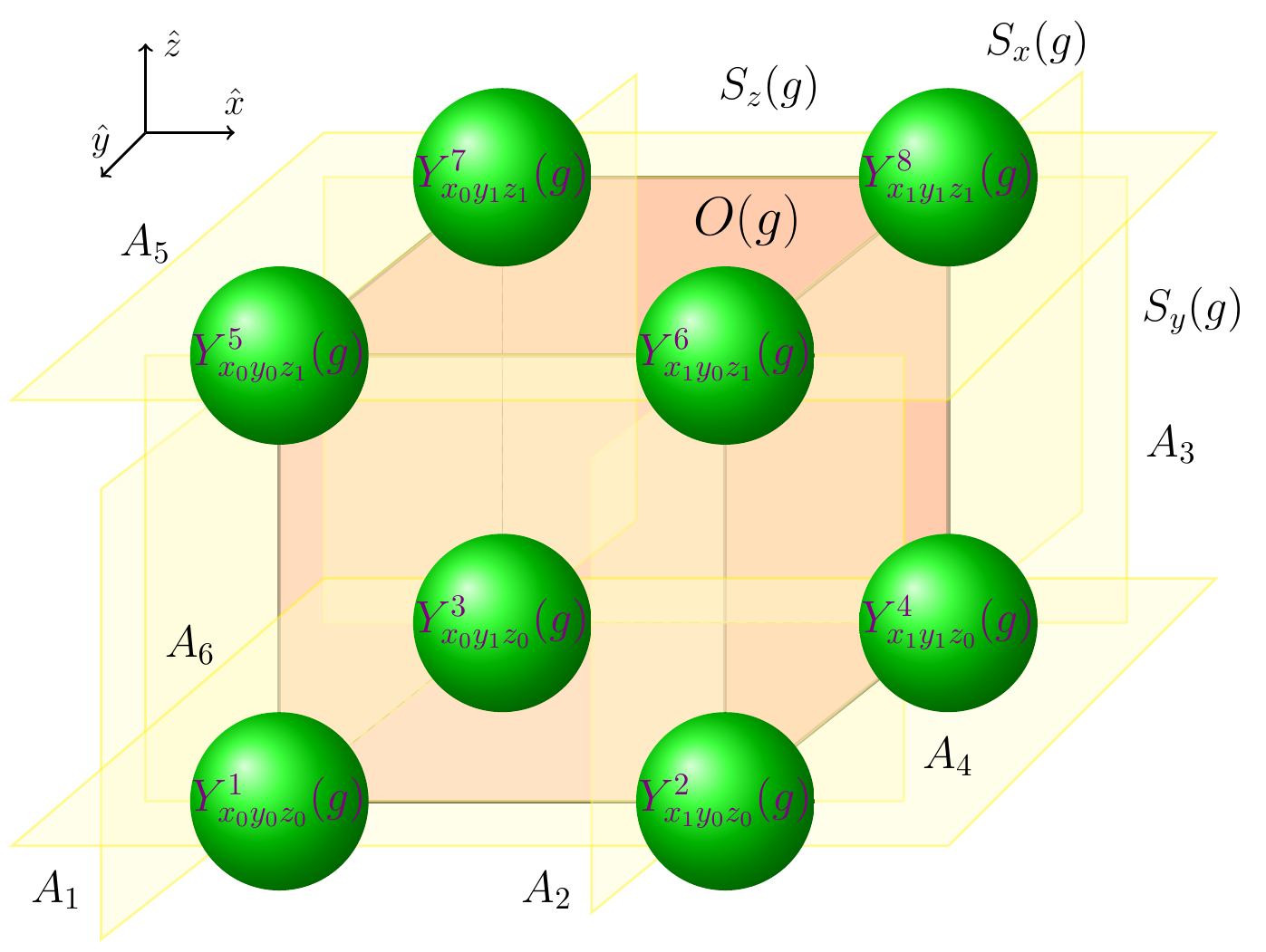}
\caption{(3+1)D lattice model with 3-foliated subsystem symmetries $U_x(g)$, $U_y(g)$ and $U_z(g)$. Yellow plates depict subsystem symmetries, and the truncated symmetry operator $O(g)$ is depicted by a red cubic that only creates 0D excitations at the corners as illustrated by green balls. $A_j$ ($j=1,2,3,4,5,6$) depicts the background gauge fields of corresponding subsystem symmetries.}
\label{3-foliated}
\end{figure}

We consider systems with homogeneous
 subsystem symmetries and derive a complete classification. For (3+1)D systems with 3-foliated homogeneous subsystem symmetries, there is an on-site symmetry group $G_s$ on each site $(x,y,z)$ acting as a unitary representation $u_{xyz}(g)$ on the local Hilbert space $\mathcal{H}_{xyz}$, while the total Hilbert space is $\mathcal{H}=\otimes_{x,y,z}\mathcal{H}_{xyz}$. The 3-foliated subsystem symmetries are defined as:
\begin{align}
\begin{aligned}
&U_x(g)=\prod\limits_{y=-\infty}^\infty\prod\limits_{z=-\infty}^\infty u_{xyz}(g)\\
&U_y(g)=\prod\limits_{x=-\infty}^\infty\prod\limits_{z=-\infty}^\infty u_{xyz}(g)\\
&U_z(g)=\prod\limits_{x=-\infty}^\infty\prod\limits_{y=-\infty}^\infty u_{xyz}(g)\\
\end{aligned}~,~~g\in G_s
\label{3-foliated truncate}
\end{align}
The geometry of the (3+1)D lattice model is illustrated in Fig. \ref{3-foliated}.

Consider a 3d system with finite extension along all three directions. A 3rd-order SSPT can potentially host nontrivial modes at the corners of the 3d cube. We can view the whole system as a 0-dimensional system with $G_s^{\times 6}$ onsite symmetry. And the symmetry action of $G_s^{\times 6}$ must be anomaly free. The consistency conditions from the anomaly-free requirement give us the classification of the higher-order SSPT states. 

Each of the 8 corners can at most be a projective representation of the group $G_s$ which are labeled by 2-cocycles $\nu_2^i\in \mathcal{H}^2(G_s, U(1))$ with $i=1, 2, ...,8$. We will consider the anomaly-free condition for the $G_s^{\times 6}$ symmetry step by step. First, considering the anomaly-free condition for the subsystem symmetry on each individual surface gives us a condition that, on each surface, the 4 corner projective representations together form a linear representation. In terms of the anomaly action, for instance, on the left $yz$ surface, this implies
\begin{equation}
     S_1[A_1]+S_3[A_1]+S_5[A_1]+S_7[A_1]=0.
     \label{cornerconstraint1}
\end{equation}
Similarly one can write down the other 5 conditions from the other 5 surfaces. It is not sufficient to nail down a pattern of the projective representation at the 8 corners with these 6 conditions. However, there is another set of conditions when considering two surfaces with intersections. For example, let us take the left $yz$ surface and the bottom $xy$ surface and turn on the subsystem symmetry gauge field $A_1$ and $A_4$. The anomaly-free condition in this case reads
\begin{align}
S_1[A_1+A_4]&+S_3[A_1+A_4]+S_5[A_1]+S_7[A_1]\nonumber\\
&+S_2[A_4]+S_4[A_4]=0.
\end{align} 
Take a special case where $A_1=-A_4$, we have 
\begin{equation}
S_5[A_1]+S_7[A_1]+S_2[-A_1]+S_4[-A_1]=0.
\label{cornerconstraint2}
\end{equation}
From Eq. (\ref{2-cocycles}), we know that the anomaly action is an even function of the gauge field in the 0+1d case. Therefore, we can ignore the sign of the gauge field in the above equation. Combining constraint in Eq. (\ref{cornerconstraint1}) and Eq. (\ref{cornerconstraint2}), the eight response actions have the following relation
\begin{align}
S_1[A]&=-S_2[A]=S_3[A]=-S_4[A]=S_5[A]\nonumber\\
&=-S_6[A]=S_7[A]=-S_8[A]=S[A]
\label{PRassign}
\end{align}

Each of the local degrees of freedom at these corners may carry anomalies of the subsystem symmetry, namely a projective representation of subsystem symmetry $G_s$ classified by $\mathcal{H}^2[G_s, U(1)]$. And the pattern of the projective representations on the 8 corners is specified by Eq. (\ref{PRassign}). Now let us consider the anomaly-free condition for this system. Suppose the background gauge fields of subsystem symmetries on the 6 surfaces of the cubic systems are $(A_{1,2},A_{3,4},A_{5,6})$ as shown in Fig. \ref{3-foliated}. Let us turn on $A_1$, $A_3$ and $A_5$ (i.e., consider $G_s^{(1)}\times G_s^{(3)}\times G_s^{(5)}$). The total anomaly action should vanish for these three gauge fields, which in this case gives us the following equation,
\begin{align}
&0=S[A_1]+S[A_3]+S[A_5]-S[A_1+A_3]\nonumber\\
&-S[A_1+A_5]-S[A_3+A_5]+S[A_1+A_3+A_5].
\label{tri-anomaly-free}
\end{align}
Equivalently, we can also rephrase the anomaly-free condition in terms of algebraic 2-cocycles: we define an ``obstruction'' map from $\mathcal{H}^2[G_s,U(1)]$ to $\mathcal{H}^2[G_s^3,U(1)]$, as
\begin{align}
f&(\nu_2)\left[(g_1,g_1',g_1''),(g_2,g_2',g_2'')\right]\nonumber\\
=&\frac{\nu_2(g_1,g_2)\nu_2(g_1',g_2')\nu_2(g_1'',g_2'')\nu_2(g_1g_1'g_1'',g_2g_2'g_2'')}{\nu_2(g_1g_1',g_2g_2')\nu_2(g_1g_1'',g_2g_2'')\nu_2(g_1'g_1'',g_2'g_2'')}
\label{tri-obstruction}
\end{align}
The anomaly-free condition of the corners [cf. Eq. (\ref{tri-anomaly-free})] requires that the (3+1)D nontrivial SSPT phases with gapless corner modes are labeled by 2-cocycles that will be mapped to a 2-coboundary of the $G_s^3$ group, $\mathcal{B}^2[G_s^3,U(1)]$, under $f$-map, otherwise we call the corresponding 2-cocycle is \textit{obstructed}. It is easy to show that considering any other combinations of gauge fields gives us the same condition. 

Next, we turn to the case with an additional global symmetry $G_g$. The anomaly response at a corner (e.g., No. 1) in the presence of both subsystem background gauge field $A_s$ and global background gauge field $A_g$ takes the form of the following:
\begin{align}
S_1[A_s]+S_1[A_s,A_g]
\end{align}
The first term is the subsystem symmetry anomaly that we have already discussed above. The second term stresses the mixed anomaly of subsystem and global symmetries, hence it can only be nontrivial when both $A_s$ and $A_g$ are nontrivial. From the same anomaly vanishing as before, we find that
\begin{align}
S_1[A_s, A_g]&=-S_2[A_s, A_g]=S_3[A_s, A_g]=-S_4[A_s, A_g]\nonumber\\
&=S_5[A_s, A_g]=-S_6[A_s, A_g]=S_7[A_s, A_g]\nonumber\\
&=-S_8[A_s, A_g]=S[A_s, A_g]
\end{align}
and
\begin{align}
0&=S[A_{s,1},A_g]+S[A_{s,3},A_g]+S[A_{s,5},A_g]\nonumber\\
&-S[A_{s,1}+A_{s,3},A_g]
-S[A_{s,1}+A_{s,5},A_g]\nonumber\\
&-S[A_{s,3}+A_{s,5},A_g]+S[A_{s,1}+A_{s,3}+A_{s,5},A_g]
\end{align}
Then we turn into the algebraic cocycle expression that rephrase $S[A_s,A_g]$ as a 2-cocycle $\nu$ in $\mathcal{H}^2[G_s\times G_g, U(1)]$. Define the obstruction function $f$ from $\nu$ to $\mathcal{H}^2[G_s^3\times G_g, U(1)]$ as
\begin{widetext}
\begin{align}
f(\nu)&[(g_1,g_1',g_1'',h_1),(g_2,g_2',g_2'',h_2)]=\frac{\nu[(g_1,h_1),(g_2,h_2)]\nu[(g_1,h_1),(g_2,h_2)]\nu[(g_1,h_1),(g_2,h_2)]\nu[(g_1g_1'g_1'',h_1),(g_2g_2'g_2'',h_2)]}{\nu[(g_1g_1',h_1),(g_2g_2',h_2)]\nu[(g_1g_1'',h_1),(g_2g_2'',h_2)]\nu[(g_1'g_1'',h_1),(g_2'g_2'',h_2)]}
\label{3-foliated anomaly free}
\end{align}
\end{widetext}
Here we have denoted the group elements of $G_s^3\times G_g$ as $(g, g', g'', h)$. The anomaly-free condition is that $f(\nu)$ corresponds to the trivial class in $\mathcal{H}^2[G_s^3\times G_g, U(1)]$.

\begin{figure}
\centering\includegraphics[width=0.46\textwidth]{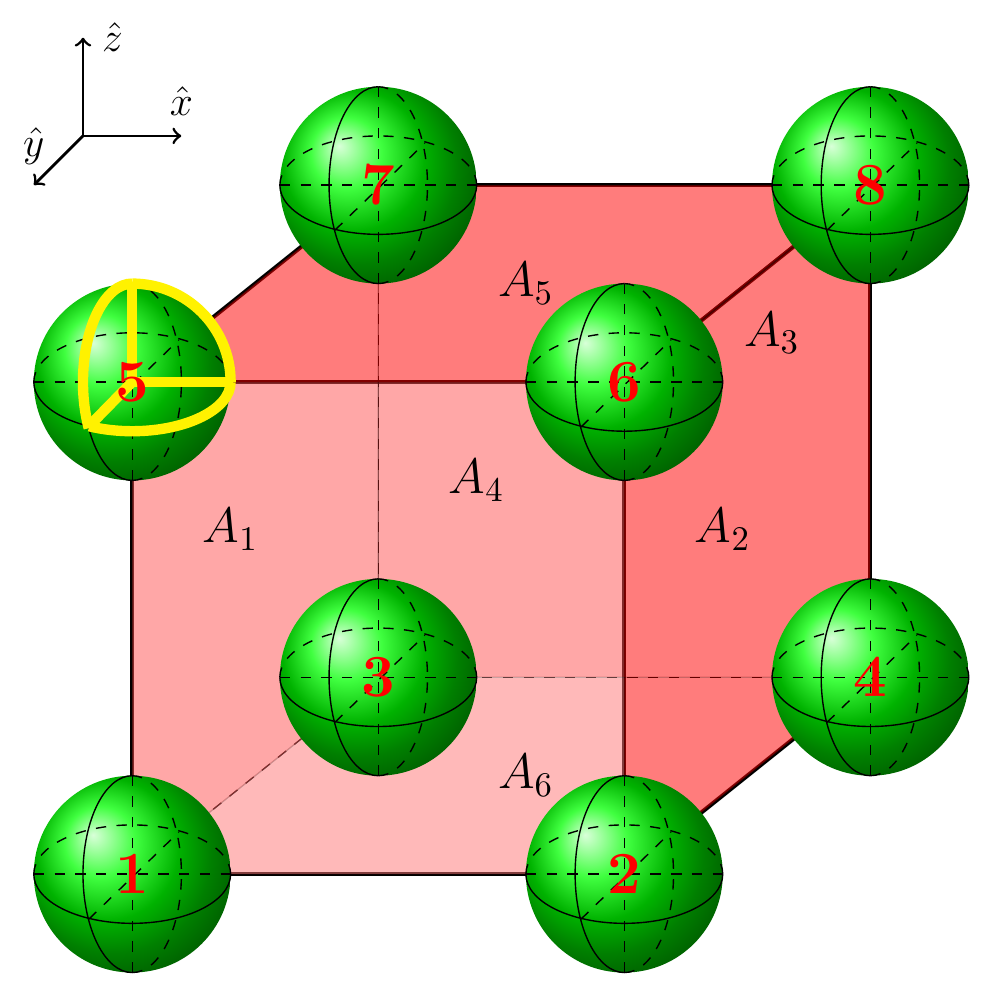}
\caption{Lattice model construction of (3+1)D SSPT phases protected by 3-foliated codimension-1 subsystem symmetries. Each green sphere depicts a lattice site which is divided into eight equal pieces (the regime enclosed by yellow segments), and there is a projective representation of $G_s$ in each of these pieces. Each red cubic stresses the eight-body interactions in the bulk. $A_1$, $A_2$, $A_3$, $A_4$, $A_5$, and $A_6$ represent the background gauge fields of the subsystem symmetries defined on the left, right, behind, front, top, and bottom surface of the red cubic, respectively.}
\label{corner}
\end{figure}

\subsection{Microscopic constructions}
We now demonstrate that all obstruction-vanishing anomalies can be realized by exact lattice model construction. Consider a (3+1)D lattice model in which each site contains four pairs of projective representations of $G_s$, each pair includes a projective representation and its inverse (see each green sphere in Fig. \ref{corner}). A cubic piece that overlaps with eight nearby sites (red cubic in Fig. \ref{corner}) also includes eight projective representations of subsystem symmetries defined on different plates. An obstruction-free 2-cocycle satisfying the Eq. (\ref{tri-obstruction}) leads to an anomaly-free red cubic in Fig. \ref{corner} automatically, hence we can always introduce some proper interaction to gap out each cubic piece in the lattice, and a fully-gapped bulk has been obtained by our construction. 

For each site on the (2+1)D surfaces, there are two pairs of dangling modes that are not included in the bulk cubic interactions, at which we can simply introduce two on-site 2-body couplings to cancel their anomalies pairwise, and a fully-gapped surface state has been obtained.

For each site on the (1+1)D hinges, there are three pairs of dangling modes that are not included in the bulk cubic interactions. Similar to the surface sites, we can introduce three on-site 2-body couplings to cancel their anomalies pairwise and obtain a fully-gapped hinge state.

Finally, for each site at the corners, there are seven dangling modes that are not included in the bulk cubic interactions, at which six of which can be gapped pairwise, similar to the arguments on the hinges. Therefore, there is a dangling 0D mode at each corner of the system with OBC, as a projective representation of the subsystem symmetry $G_s$. 


\subsection{Example\label{Sec.tri-example}}
In this subsection, we present a spin model of a (3+1)D nontrivial SSPT phase protected by 3-foliated subsystem symmetry $G_s=\mathbb{Z}_2\times\mathbb{Z}_2$, without global symmetry, by commuting-projector Hamiltonians. 

Take the ultraviolet (UV) limit of the truncated operator (\ref{3-foliated truncate}) that is defined in a specific cubic (see Fig. \ref{corner}), there are eight spin-1/2 degrees of freedom as projective representations of the $\mathbb{Z}_2\times\mathbb{Z}_2$ group, on each site (green sphere in Fig. \ref{corner}), while each red cubic also includes eight spin-1/2 degrees of freedom $\bs{S}_j$ ($j=1,\cdot\cdot\cdot,8$) from nearby lattice sites. We present the following commuting-projector lattice Hamiltonian:
\begin{align}
H_x=&\sum\limits_{\bs{R}}\left(S_{\bs{R},1}^xS_{\bs{R},2}^xS_{\bs{R},3}^xS_{\bs{R},4}^x+S_{\bs{R},1}^xS_{\bs{R},4}^xS_{\bs{R},5}^xS_{\bs{R},8}^x\right.\nonumber\\
&\left.+S_{\bs{R},1}^xS_{\bs{R},2}^xS_{\bs{R},5}^xS_{\bs{R},6}^x+S_{\bs{R},2}^xS_{\bs{R},4}^xS_{\bs{R},5}^xS_{\bs{R},7}^x\right)
\end{align}
\begin{align}
H_z=&\sum\limits_{\bs{R}}\left(S_{\bs{R},1}^zS_{\bs{R},2}^zS_{\bs{R},3}^zS_{\bs{R},4}^z+S_{\bs{R},1}^zS_{\bs{R},4}^zS_{\bs{R},5}^zS_{\bs{R},8}^z\right.\nonumber\\
&\left.+S_{\bs{R},1}^zS_{\bs{R},2}^zS_{\bs{R},5}^zS_{\bs{R},6}^z+S_{\bs{R},2}^zS_{\bs{R},4}^zS_{\bs{R},5}^zS_{\bs{R},7}^z\right)
\end{align}
It is straightforward to check that the total Hamiltonian $H_x+H_z$ respects the $\mathbb{Z}_2\times\mathbb{Z}_2$ subsystem symmetry, and provides a non-degenerate ground state. By definition, this spin model will give a dangling spin-1/2 degree of freedom at each corner of the system: for the lattice site on the (2+1)D surface, there are four dangling spin-1/2 degrees of freedom that are not included in the bulk cubic interactions $H_x+H_z$, with on-site $\mathbb{Z}_2\times\mathbb{Z}_2$ symmetry. We can simply introduce a pair of Heisenberg interaction $\bs{S}\cdot\bs{S}'$ on each of these sites to fully gap them out. Similar to the sites on the (1+1)D hinges and corners, there will be a dangling spin-1/2 degree of freedom on each corner of the open system. 

On the other hand, a spin-1/2 degree of freedom corresponds to a projective representation of the $\mathbb{Z}_2\times\mathbb{Z}_2$ group, which is labeled by a 2-cocycle $\nu_2\in\mathcal{H}^2[\mathbb{Z}_2\times\mathbb{Z}_2,U(1)]$, with the following explicit expression:
\begin{align}
\nu_2(a,b)=(-1)^{a_1b_2}
\label{2-cocycle of Z2*Z2}
\end{align}
where $a=(a_1,a_2)$, $b=(b_1,b_2)$ and $a_{1,2},b_{1,2}\in\mathbb{Z}_2$. Substitute this explicit form of 2-cocycle into Eq. (\ref{tri-obstruction}), and we explicitly find that the anomaly-free condition (\ref{3-foliated anomaly free}) is satisfied automatically. Hence the nontrivial 2-cocycle (\ref{2-cocycle of Z2*Z2}) of $\mathbb{Z}_2\times\mathbb{Z}_2$ labels a nontrivial third-order SSPT phase protected by 3-foliated $\mathbb{Z}_2\times\mathbb{Z}_2$ subsystem symmetry, which is consistent with above lattice spin model construction.

\section{Summary\label{Sec.summary}}
In this work, we systematically construct and classify interacting fractonic higher-order topological phases in (2+1)D and (3+1)D protected by 2-foliated and 3-foliated subsystem symmetries and global symmetries. 

For 2-foliated homogeneous subsystem symmetry $G_s$ and a global symmetry $G
_g$ in ($d$+1)D systems, a possible nontrivial SSPT phase is labeled by a $d$-cocycle $\nu_d\in\mathcal{H}^d[G_s\times G_g, U(1)]$ satisfying an anomaly free conditions (\ref{2-foliated anomaly free1}) for $G_g$ trivial and (\ref{2-foliated anomaly free2}) for nontrivial global symmetry $G_g$. We prove that there is no nontrivial 2-foliated SSPT phase in (2+1)D systems without a proper global symmetry for Abelian subsystem symmetry $G_s$. We present several explicit examples of of 2-foliated SSPT phases using coupled-wire model constructions. In (2+1)D, we construct a spin model on a square lattice with ring exchange in each plaquette that realizes a higher-order SSPT protected by $G_s=\mathbb{Z}_2$ and $G_g=\mathbb{Z}_2$. In (3+1)D, we first consider a case with $G_s=\mathbb{Z}_2$ and no global symmetry. A wire construction lattice model for a higher-order SSPT using the edge mode of Levin-Gu state is constructed. Subsequently, we consider $G_s=\mathbb{Z}_2$ and ${G}_g=\mathbb{Z}_2^{\mathrm{T}}$ and build the coupled-wire model from the Luttinger liquid carrying the mixed anomaly of $\mathbb{Z}_2$ and $\mathbb{Z}_2^{\mathrm{T}}$. Finally, we consider a fermionic example with $G_s=\mathbb{Z}_2^f$ and $G_g=\mathbb{Z}_2$, while the coupled-wire model is constructed from the Luttinger liquid as the edge mode of fermionic Levin-Gu state \cite{Gu-Levin}. 

For 3-foliated homogeneous subsystem symmetry $G_s$ and a global symmetry $G_g$ in (3+1)D systems, a possible nontrivial SSPT phase is labeled by a 2-cocycle in $\mathcal{H}^2[G_s\times G_g, U(1)]$ satisfying the anomaly-free condition (\ref{tri-obstruction}). We provide an example of (3+1)D SSPT phase with a spin-1/2 degree of freedom as the concrete third-order topological corner mode, protected by subsystem symmetry $G_s=\mathbb{Z}_2\times\mathbb{Z}_2$, by exact solvable lattice model construction in Sec. \ref{Sec.tri-example}.

Furthermore, in Appendix \ref{Inhomo1} and \ref{Inhomo2}, we prove that despite the dimensionality of the system with any global symmetry $G_g$, the inhomogeneous subsystem symmetry will always give a trivial SSPT phase with a layered structure, which will always be trivialized by adhering to lower-dimensional SPT phases protected by subsystem symmetry $G_s$ acting internally, to the surface or hinge of the open boundary of the system. 

Our construction and classification of SSPT phases based on the anomaly-free condition are well-defined not only in the bosonic systems but can also be generalized to interacting fermionic systems, and the fermionic higher-order SSPT phases might also be characterized by lower-dimensional group super-cohomological cocycles. Furthermore, with various higher-order SSPT phases, we can investigate the possible fracton phases obtained from the higher-order SSPT phases constructed in this work by gauging the subsystem symmetry \cite{Shirley_2019}. 

\begin{acknowledgements}
We thank Ruochen Ma for the helpful discussions. JHZ thanks the hospitality of Liujun Zou and Chong Wang at the Perimeter Institute for Theoretical Physics, where part of this work is finished. JHZ and ZB are supported by startup funds from The Pennsylvania State University. The work of MC was supported in part by NSF under award number DMR-1846109.
\end{acknowledgements}

\appendix

\section{Comment on 1-foliated systems and higher-order SSPT}
\label{1foliated}
In this section, we comment on 1-foliated systems and higher-order SSPT. We argue that there is no nontrivial higher-order SPT phase protected by 1-foliated subsystem symmetries even with the help of a global symmetry. In order to achieve a higher-order SSPT phase, we need to have some modes that are anomalous appearing on the corners/hinges. We will argue that, with the 1-foliated structure, such anomalous corner modes can be canceled by attaching a lower dimensional SPT on the boundary of the system. We will focus on the case with only subsystem symmetry first. 

Let us consider the 2d case. For reference, we can look at Fig. \ref{2dSSym}. But in the current case, we only have subsystem symmetry along the y-direction and no subsystem symmetry along $x$. In another word, we only have $A_1$ and $A_2$ background gauge fields available. Consider a finite 2d system, if there are nontrivial modes (in this case just projective representations) on the 4 corners of the system, all of them combined together must be anomaly-free in order to have a well-defined 2d system. Analogous arguments similar to the main text give us anomaly-free conditions where the two projective representations on the left edge must be opposite to each other and the same for the right edge. The crucial difference from the 2-foliation case is that there is no subsystem symmetry connecting the left and the right edge, i.e., no additional anomaly-free conditions. Now we can see that we can indeed attach a 1d SPT with $G_s$ symmetry, whose boundary precisely corresponds to the projective representations on the corner, to cancel the corner modes. This is not possible for the 2-foliated case because the corner modes carry additional anomalies of the subsystem symmetry from the other foliation direction. 

In 3d, the case for third-order SSPT, namely an SSPT with nontrivial modes at the corners of cubic systems, is very similar to the 2d case we argued above. The same argument for 2d also shows that there is no nontrivial 3d order-2 SSPT. We will not elaborate on it here.

In 3d, the case for second-order SSPT is interesting. Suppose we have a 1-foliation subsystem symmetry along the $z$-direction, namely in each $xy$-plane there is a symmetry group $G_s$. Consider a finite cube, there are several possibilities for second-order SSPT. We can have hinge modes along the $x$ or $y$ directions. In this case, the anomaly-free condition is simply that anomalies of the two modes on the same $xy$-plane must cancel each other. With this, we can just attach a 2d $G_s$ SPT on the 2d surface to cancel the hinge mode. Again we emphasize that this attachment is not possible in the 2-foliated case because the hinge mode also carries an anomaly of the subsystem symmetry from the orthogonal direction. A more subtle question is whether we can have nontrivial higher-order SSPT where the hinge mode runs along the $z$-direction. We argue that there is no such nontrivial SPT either. Suppose we have 4 anomalous hinge modes running along $z$-direction on the 4 hinges of the system. A single hinge mode will carry the anomaly of $G_s^{\times L_z}$ where $L_z$ is the length of $z$-direction. In this case, because each $G_s$ only acts on a single unit cell, the most general anomaly pattern is just given by a series of projective representations of $G_s$ along the chain. In order for the whole system to be anomaly-free, on each $xy$-plane the 4 projective representations from the 4 hinges must cancel together. This tells us in each $xy$-plane we can attach 1d SPTs on the 4 boundaries to cancel all the projective representations on the corners. Again this procedure can be done because we have no additional symmetry constraints from other directions. 

Now we consider the effect of the addition of a global symmetry $G_g$. It is easy to see that, if the corner/hinge modes carry only $G_g$ anomaly, one can always attach a surface $G_g$ SPT to cancel that anomaly without doing any bulk modifications. If the corner/hinge modes carry mixed anomaly of $G_s$ and $G_g$, one can show that the anomaly-free condition is still the anomalous modes from the same foliation edge/surface cancel together. This in essence means we can attach a lower dimensional SPT on the edge/surface to cancel all the anomalous corners/hinges.

\section{Inhomogeneous cases are always trivial}
\label{Inhomo1}


In order to investigate the higher-order SSPT phases, we suppose the symmetry operators only have nontrivial effects at the corner of the regime. Hence the symmetry operator (\ref{2-foliated inhomogeneous})  still creates four local excitations at the codimension-2 corners of the regime.

For bosonic systems in $D\leq2$ with unitary symmetry $G_s^x\times G_s^y$, each anomaly class is uniquely determined by the group cohomology class $[\nu]\in\mathcal{H}^{D+2}[G_s^x\times G_s^y, U(1)]$. This cohomology class can be rephrased by the K\"unneth formula as
\begin{align}
\mathcal{H}^{D+2}[G_s^x\times G_s^y, U(1)]=\prod\limits_{p=0}^{D+2}\mathcal{H}^{D+2-p}\left(G_s^x,\mathcal{H}^p[G_x^y, U(1)]\right)
\label{Kunneth formula}
\end{align}
i.e., the first and last term of Eq. (\ref{Kunneth formula}) depict the unique anomaly of $G_s^x$ and $G_s^y$, respectively; all other terms depict the mixed anomaly of $G_s^x$ and $G_s^y$. The action of anomaly is formally written as
\begin{align}
S_{\mathrm{anomaly}}^p[A_x,A_y]=\int_{\hat{\mathcal{L}}_y}A_x^*\nu_x^p,~p=0,\cdot\cdot\cdot,D+2
\label{2-foliated anomaly}
\end{align}
where $A_x$ is the background gauge field of $G_s^x$ that we view it as a mapping from the spacetime manifold $M_p$ as a sub-manifold of $M_{D+2}$, to the classifying space $BG_{s}^x$, $A_x^*\nu_x^p$ is the pullback of $\mathcal{H}^p(BG_s^x,\mathbb{R}/\mathbb{Z})$, while $\hat{\mathcal{L}}_y$ is the Poincar\'e dual of the cocycle $\mathcal{L}_y=A_y^*\nu_y^{D+2-p}$ with respect to the spacetime $M_{D+2}$. Physically, $\hat{\mathcal{L}_y}$ stresses the codimension-$p$ domain wall of $G_s^y$ symmetry, and the anomaly action (\ref{2-foliated anomaly}) represents the topological response theory of a $p$-dimensional $G_s^x$-SPT phase on the codimension-$p$ domain wall of $G_s^y$. 

For the symmetry operator in Fig. \ref{2dSSym}, the background gauge fields $A_{1,2,3,4}$ correspond to the symmetry group $(G_s^x\times G_s^y)^2$. Now we turn on the background gauge field $A_1$ only. Both TL and BL corner theories are coupled to $A_1$, so the anomaly-free condition implies
\begin{align}
S_{\mathrm{BL}}[A_1]+S_{\mathrm{TL}}[A_1]=0
\label{2-foliated anomaly1}
\end{align}
Similarly, we have
\begin{align}
\begin{aligned}
&S_{\mathrm{BR}}[A_2]+S_{\mathrm{TR}}[A_2]=0\\
&S_{\mathrm{TL}}[A_3]+S_{\mathrm{TR}}[A_3]=0\\
&S_{\mathrm{BL}}[A_4]+S_{\mathrm{BR}}[A_4]=0
\end{aligned}
\label{2-foliated anomaly2}
\end{align}

We turn on the background gauge fields $A_1$ and $A_3$ (see Fig. \ref{2dSSym}), and the anomaly carried by the TL corner should be:
\begin{align}\sum\limits_{p=0}^{d+1}S_{\mathrm{TL}}^p[A_1,A_3]=\sum\limits_{p=0}^{d+1}\int_{\hat{\mathcal{L}_3}}A_1^*\nu_x^p
\label{anomaly A1 and A3}
\end{align}
The anomaly carried by TR and BL corner are $S_{\mathrm{TR}}[A_3]$ and $S_{\mathrm{BL}}[A_1]$, respectively. We notice that only the TL corner carries the mixed anomaly of $A_1$ and $A_3$ [$p\ne0$ in Eq. (\ref{anomaly A1 and A3})], hence the anomaly-free condition requires that all mixed anomalies at the TL corner vanish. Equivalently, the anomaly carried by a corner (\ref{2-foliated anomaly}) should reduce to the sum of a unique anomaly of vertical and horizontal subsystem symmetries as
\begin{align}
S_{\mathrm{anomaly}}[A_x, A_y]=S_{\mathrm{anomaly}}[A_x]+S_{\mathrm{anomaly}}[A_y]
\end{align}
Hence the anomaly-free conditions are exactly highlighted in Eqs. (\ref{2-foliated anomaly1}) and (\ref{2-foliated anomaly2}) and all mixed anomalies of subsystem symmetries should vanish. 

Then we consider a codimension-1 SPT phase protected by $G_s^y$ on the system's left edge, with the enlarged truncated symmetry operator applying to the whole system. This SPT phase leaves codimension-2 modes at the TL and BL corners, carrying the anomalies of the background gauge field $A_1$ phrased by the actions $S_{\mathrm{BL}}'[A_1]$ and $S_{\mathrm{TL}}'[A_1]$, satisfying
\begin{align}
S_{\mathrm{BL}}'[A_1]+S_{\mathrm{TL}}'[A_1]=0
\end{align}
We simply choose $S_{\mathrm{BL}}'[A_1]=-S_{\mathrm{BL}}[A_1]$ and $S_{\mathrm{TL}}'[A_1]=-S_{\mathrm{TL}}[A_1]$, and the bulk anomaly of the background gauge field $A_1$ is canceled by this codimension-1 SPT phase on the left boundary. Similarly, the bulk anomaly of the background gauge field $A_3$ can be canceled by an additional codimension-1 SPT phase protected by $G_s^x$ symmetry on the top boundary. As a consequence, there is no higher-order SSPT phase for 2-foliated inhomogeneous subsystem symmetry.

\section{Inhomogeneous subsystem symmetries in 3D}
\label{Inhomo2}
For inhomogeneous subsystem symmetries, there are several possible scenarios: 
\begin{enumerate}[1.]
\item The subsystem symmetries in all three directions act respectively that can be different. We denote the subsystem symmetries defined in $yz$-plane, $xz$-plane and $xy$-plane as $G_x$, $G_y$ and $G_z$, respectively. 
\item The subsystem symmetries in two of three directions act identically, and the other subsystem symmetry acts separately. We denote these subsystem symmetries by two groups $G_s^1$ and $G_s^2$, respectively.
\end{enumerate}
For the first scenario, the 3-foliated subsystem symmetries should be redefined as:
\begin{align}
\begin{aligned}
&U_x(g_x)=\prod\limits_{y=-\infty}^\infty\prod\limits_{z=-\infty}^\infty u_{xyz}^x(g_x)\\
&U_y(g_y)=\prod\limits_{x=-\infty}^\infty\prod\limits_{z=-\infty}^\infty u_{xyz}^y(g_y)\\
&U_z(g_z)=\prod\limits_{x=-\infty}^\infty\prod\limits_{y=-\infty}^\infty u_{xyz}^z(g_z)
\end{aligned}~,~\left\{
\begin{aligned}
&g_x\in G_x\\
&g_y\in G_y\\
&g_z\in G_z
\end{aligned}
\right.
\label{3-foliated truncated}
\end{align}
where $\left(u_{xyz}^x(g_x),u_{xyz}^y(g_y),u_{xyz}^z(g_z)\right)$ are linear representations of the groups $(G_x,G_y,G_z)$, respectively. The local degrees of freedom at the corners should be labeled by the projective representations of $G_x\times G_y\times G_z$ group, classified by $\mathcal{H}^2[G_x\times G_y\times G_z, U(1)]$. The action of anomaly is formally written as
\begin{align}
S_{\mathrm{anomaly}}^p[A_i, A_j]=\int_{\hat{\mathcal{L}}_j}A_i^*\nu_i^p,~p=0,1,2
\end{align}
where $A_i$ and $A_j$ are background gauge fields of $G_s^i$ and $G_s^j$ ($i,j=x,y,z$ and $i\ne j$) that we view them as maps from the spacetime manifolds $M_p$ and $M_{2-p}$ as the sub-manifolds of $M_2$ to the classifying spaces $BG_s^i$ and $BG_s^j$, and $A_{i,j}^*\nu_{i.j}^p$ is the pullback of $\mathcal{H}^p(BG_{s}^{i,j},\mathbb{R}/\mathbb{Z})$, while $\hat{\mathcal{L}}_j$ is the Poincar\'e dual of the cocycle $\mathcal{L}_j=A_j^*\nu_j^{2-p}$ with respect to the spacetime $M_2$. 

For the truncated symmetry operator in Fig. \ref{3-foliated}, the background gauge fields $A_j$ ($j=1,\cdot\cdot\cdot,6$) correspond to the symmetry group $(G_s^x\times G_s^y\times G_s^z)^2$. We first turn on the background gauge field $A_1$ only, the anomaly-free condition implies the following constraints on the anomaly action as
\begin{align}
S_1[A_1]+S_3[A_1]+S_5[A_1]+S_7[A_1]=0
\end{align}
and other five similar constraints from the other five surfaces. 

Then we turn on the background gauge fields $A_1$ and $A_4$, and the total anomaly carried by the corner-1/-4 is phrased by the following action
\begin{align}
\sum\limits_{p=0}^2S_j^p[A_1,A_4]=\sum\limits_{p=0}^2\int_{\hat{\mathcal{L}}_4}A_1^*\nu_x^p,~j=1,4.
\end{align}
The mixed anomalies carried by corner-1 and corner-4 satisfy the following constraint to ensure the anomaly-free condition as
\begin{align}
S_1^1[A_1,A_4]=-S_4^1[A_1,A_4]
\end{align}
and similar for all other mixed anomalies. Therefore, we always have some way to find an anomaly-free bulk lattice model. 

Then we consider the possible trivializations of the constructed model of 3-foliated inhomogeneous SSPT phases. For the corner mode carrying a unique anomaly of the subsystem symmetry along a specific direction, for instance, in $yz$-plane with background gauge field $A_1$, which is labeled by a 2-cocycle $\nu_2\in\mathcal{H}^2[G_x, U(1)]$ at the corner-7, consider the (1+1)D hinge of left and behind surfaces, at which the subsystem symmetries $G_x$ and $G_y$ act as global symmetries. Consider a (1+1)D SPT phase protected by $G_x$ symmetry while $G_y$ and $G_z$ act on it trivially, we choose $-\nu_2\in\mathcal{H}^2[G_x,U(1)]$ to characterize this (1+1)D SPT phase, whose dangling edge mode at corner-7 will annihilate the dangling gapless mode from bulk lattice model construction, i.e., the corresponding bulk state is \textit{trivialized}. Similar to $G_y$ and $G_z$, all corner modes carrying the unique anomaly of $G_{x,y,z}$ can be trivialized by adhering a (1+1)D SPT phase to the hinge of the cubic.

For the corner mode carrying a mixed anomaly of the subsystem symmetries along two directions. Without loss of generality, we consider the (0+1)D dangling mode carrying the mixed anomaly of $G_x$ and $G_y$, characterized by a cup product of two 1-cocycles:
\begin{align}
\nu_1^x(g_x)\cup\nu_1^y(g_y),~\left\{
\begin{aligned}
&\nu_1^x(g_x)\in\mathcal{H}^1[G_x,U(1)]\\
&\nu_1^y(g_y)\in\mathcal{H}^1[G_y,U(1)]
\end{aligned}
\right.
\end{align}
Repeatedly consider the (1+1)D hinge of left and behind surfaces, at which $G_x$ and $G_y$ act as global symmetries. Consider a (1+1)D SPT phase protected by $G_x\times G_y$ while $G_z$ acts on it trivially, we choose $-\nu_1^x\cup\nu_1^y\in\mathcal{H}^1[G_x,\mathcal{H}^1[G_y,U(1)]]$ to characterize this (1+1)D SPT phase, whose dangling edge mode at corner-7 will annihilate the dangling gapless mode from bulk lattice model construction, i.e., the corresponding bulk state is also trivialized. Similar for all other mixed anomalies carried by corner modes. 

We conclude that for the scenario of which the subsystem symmetries in all three directions act respectively, the classification of third-order SSPT phases should be trivial because all possible bulk lattice model constructions are trivialized by adhering a (1+1)D SPT phase to the hinges of the system.

For the second scenario that the subsystem symmetries in two of three directions act identically, while the subsystem symmetry in the other direction act separately, the 3-foliated subsystem symmetries should be redefined as (without loss of generality, we suppose the subsystem symmetry on the $xz$-plane and $yz$-plane is $G_s^1$, while the subsystem symmetry on the $xy$-plane is $G_s^2$):
\begin{align}
\begin{aligned}
&U_x(g_1)=\prod\limits_{y=-\infty}^\infty\prod\limits_{z=-\infty}^\infty u_{xyz}^1(g_1)\\
&U_y(g_1)=\prod\limits_{x=-\infty}^\infty\prod\limits_{z=-\infty}^\infty u_{xyz}^1(g_1)\\
&U_z(g_2)=\prod\limits_{x=-\infty}^\infty\prod\limits_{y=-\infty}^\infty u_{xyz}^2(g_2)
\end{aligned}~,~\left\{
\begin{aligned}
&g_1\in G_s^1\\
&g_2\in G_s^2
\end{aligned}
\right.
\end{align}
where $u_{xyz}^1(g_1)$ and $u_{xyz}^2(g_2)$ are linear representations of the groups $G_s^1$ and $G_s^2$, respectively. The local degrees of freedom at the corners are projective representations of the group $G_s^1\times G_s^2$, classified by:
\begin{align}
&\mathcal{H}^2[G_s^1\times G_s^2,U(1)]=\mathcal{H}^2[G_s^1,U(1)]\nonumber\\
&\times\mathcal{H}^2[G_s^2,U(1)]\times\mathcal{H}^1[G_s^1,\mathcal{H}^1[G_s^2,U(1)]],
\end{align}
in terms of background gauge fields, the action of the anomaly of $G_s^1\times G_s^2$ is formally written as
\begin{align}
S_{\mathrm{anomaly}}^p[A_{1,s},A_{2,s}]=\int_{\hat{\mathcal{L}}_2}A_{1,s}^*\nu_1^p,~p=0,1,2
\end{align}
where $A_1$ and $A_2$ are background gauge fields of $G_s^1$ and $G_s^2$ that we view as maps from the spacetime manifolds $M_p$ and $M_{2-p}$ as the sub-manifolds of $M_2$ to the classifying spaces $BG_s^1$ and $BG_s^2$, respectively; $A_{j,s}^*\nu_j^p$ ($j=1,2$) is the pullback of $\mathcal{H}^p(BG_s^j, \mathbb{R}/\mathbb{Z})$, while $\hat{\mathcal{L}}_2$ is the Poincar\'e dual of the cocycle $\mathcal{L}_2=A_{2,s}^*\nu_2^{2-p}$ with respect to the spacetime $M_2$. 

Following the similar arguments with above $G_s^x\times G_s^y\times G_s^z$ cases, we conclude that we can always find some fully-gapped, anomaly-free bulk lattice model. Then we focus on some possible trivializations by sticking some lower-dimensional SPT phases on the edge/hinge. 

Firstly, for corner-7 carrying a projective representation $\nu_2^2\in\mathcal{H}^2[G_s^2, U(1)]$, we focus on the hinge of the top and behind surfaces at which the subsystem symmetry $G_s^2$ acts as a global symmetry: Consider a (1+1)D SPT phase protected by $G_s^2$ adhered to this hinge, while the subsystem symmetry $G_s^1$ acts on this (1+1)D SPT state trivially. We choose $-\nu_2^2\in\mathcal{H}^2[G_s^2,U(1)]$ to characterize this adhered (1+1)D SPT phase, whose dangling edge mode at the corner-7 will annihilate the dangling gapless mode from bulk lattice model construction, i.e., the corresponding bulk state is \textit{trivialized}.

Subsequently, for corner-7 carrying a projective representation $\nu_2^1\in\mathcal{H}^2[G_s^1,U(1)]$, we focus on the hinge of the left and behind surfaces at which the subsystem symmetry $G_s^1$ acts as a global symmetry: Consider a (1+1)D SPT phase protected by $G_s^1$ adhered to this hinge, while the subsystem symmetry $G_s^2$ acts on this (1+1)D SPT state trivially. We choose $-\nu_2^1\in\mathcal{H}^2[G_s^1,U(1)]$ to characterize this adhered (1+1)D SPT phase, whose dangling edge mode at the corner-7 will annihilate the dangling gapless mode from bulk lattice model construction, i.e., the corresponding bulk state is \textit{trivialized}.

Finally, for corner-7 carrying a projective representation with the mixed anomaly of $G_s^1$ and $G_s^2$, we repeatedly focus on the hinge of the top and behind surfaces at which the subsystem symmetries defined on the $xy$-plane and $xz$-plane act as global symmetries: consider a (1+1)D SPT phase protected by $G_s^1\times G_s^2$ (from the subsystem symmetries defined on the $xy$-plane and $xz$-plane) adhered to this hinge, while the subsystem symmetry defined on the $yz$-plane acts on this (1+1)D SPT state trivially. We choose $-\nu_1^1\cup\nu_1^2\in\mathcal{H}^1[G_s^1,\mathcal{H}^1[G_s^2,U(1)]]$ to characterize this adhered (1+1)D SPT phase, whose dangling edge mode at the corner-7 will annihilate the dangling gapless mode from bulk lattice model construction, i.e., the corresponding bulk state is also \textit{trivialized}. 

Now we have rigorously proved that for 3-foliated inhomogeneous subsystem symmetries in (3+1)D systems, there is no nontrivial SSPT phase with a third-order topological corner state.

\providecommand{\noopsort}[1]{}\providecommand{\singleletter}[1]{#1}%
%


\end{document}